 \documentclass[final,1p,times]{elsarticle}
\usepackage{amssymb}
\usepackage{color,soul}
\usepackage[hang,bottom]{footmisc}
\usepackage{amsmath} 
\usepackage{framed} % Framing content
\usepackage{multicol} % Multiple columns environment
\usepackage{nomencl} % Nomenclature package
\makenomenclature
\renewcommand*\nompreamble{\begin{multicols}{2}}
\renewcommand*\nompostamble{\end{multicols}}
\usepackage{siunitx} 
\usepackage{color}
\usepackage{tabularray}
\usepackage{array}
\usepackage{booktabs}
\usepackage{longtable}
\usepackage{subcaption}
\usepackage{url}

 \usepackage{array}
 \usepackage{pdflscape}
 \usepackage{longtable}
 \usepackage{booktabs}
\usepackage{hyperref}

\usepackage[utf8]{inputenc}
\journal{arXiv}

\begin{document}

\begin{frontmatter}

%% Title, authors and addresses

%% use the tnoteref command within \title for footnotes;
%% use the tnotetext command for theassociated footnote;
%% use the fnref command within \author or \address for footnotes;
%% use the fntext command for theassociated footnote;
%% use the corref command within \author for corresponding author footnotes;
%% use the cortext command for theassociated footnote;
%% use the ead command for the email address,
%% and the form \ead[url] for the home page:
%% \title{Title\tnoteref{label1}}
%% \tnotetext[label1]{}
%% \author{Name\corref{cor1}\fnref{label2}}
%% \ead{email address}
%% \ead[url]{home page}
%% \fntext[label2]{}
%% \cortext[cor1]{}
%% \affiliation{organization={},
%%             addressline={},
%%             city={},
%%             postcode={},
%%             state={},
%%             country={}}
%% \fntext[label3]{}

\title{AI and Quantum Computing in Binary Photocatalytic Hydrogen Production}

\author[ftkkp]{Dennis Delali Kwesi Wayo} 
\ead{dennis.wayo@nu.edu.kz}

\author[dcam]{Leonardo Goliatt} 
\ead{leonardo.goliatt@ufjf.br}

\author[dcn]{Darvish Ganji} 
\ead{ganji_md@yahoo.com}

\affiliation[ftkkp]{organization={Faculty of Chemical and Process Engineering Technology, Universiti Malaysia Pahang Al-Sultan Abdullah},
            city={Kuantan},
            postcode={26300}, 
            country={Malaysia}}

\affiliation[dcam]{organization={Department of Computational and Applied Mechanics, Federal University of Juiz de Fora},
            city={Juiz de Fora},
            postcode={36036-900}, 
            country={Brazil}}

\affiliation[dcn]{organization={Division of Carbon Neutrality and Digitalization, Korea Institute of Ceramic Engineering and Technology (KICET)},
            city={Jinju},
            postcode={52851}, 
            country={Republic of Korea}}

%%%%%%%%%%%%%%%%%%%%%%%%%%%%%%%%%%%%%%%%%%%%%%%%%%%%%%%%%%%%%%%%%%%%%%%%

\begin{abstract}
Photocatalytic water splitting has emerged as a sustainable pathway for hydrogen production, leveraging sunlight to drive chemical reactions. This review explores the integration of density functional theory (DFT) with machine learning (ML) to accelerate the discovery, optimization, and design of photocatalysts. DFT provides quantum-mechanical insights into electronic structures and reaction mechanisms, while ML algorithms enable high-throughput analysis of material properties, prediction of catalytic performance, and inverse design. This paper emphasizes advancements in binary photocatalytic systems, highlighting materials like TiO\(_2\), BiVO\(_4\), and g-C\(_3\)N\(_4\) as well as novel heterojunctions and co-catalysts that improve light absorption and charge separation efficiency. Key breakthroughs include the use of ML architectures such as Random Forests, Support Vector Regression, and Neural Networks, trained on experimental and computational datasets to optimize band gaps, surface reactions, and hydrogen evolution rates. Emerging techniques like Quantum Machine Learning (QML) and generative models (GANs, VAEs) demonstrate the potential to explore hypothetical materials and enhance computational efficiency. The review also highlights advanced light sources, such as tunable LEDs and solar simulators, for experimental validation of photocatalytic systems. Challenges related to data standardization, scalability, and interpretability are addressed, proposing collaborative frameworks and open-access repositories to democratize DFT-AI tools. By bridging experimental and computational methodologies, this synergistic approach offers transformative potential for achieving scalable, cost-effective hydrogen production, paving the way for sustainable energy solutions.

\end{abstract}

\vfill


\end{frontmatter}

%--------------------------------------------------------
\section{Introduction}
%--------------------------------------------------------

Carbon saturation in the atmosphere has created an immerse climate change \cite{karl2009global, wheeler2013climate, sesana2021climate, abbass2022review, corwin2021climate} emanating from heavy industrial consumption of fossil fuel, which accounts to 76\% of greenhouse gas emissions (GHG). There are currently about 265 carbon capture project sites and about 40 of them are operational \cite{Statista2024}, considering about 37 billion metric ton of globally carbon emissions, these carbon capture techniques and the number of available operational sites aren’t as promising as the recent advancements made in renewable energies. Nevertheless, as scientists and engineers strive to eliminate the use of hydrocarbons for both pre and post combustion processes, hydrogen has emerged as a promising alternative source of clean energy.

The rising demand of hydrogen is intended to improve and develop the human index. For this reason, many techniques for the production of hydrogen \cite{dawood2020hydrogen, abdin2020hydrogen, yue2021hydrogen} have been developed with concerns to reducing the energy intensity to hydrogen production without emitting GHG and at a low cost. Some of these hydrogen production resources which are considered safe are from biomass, solar, and wind with several evolving semiconducting catalytic materials. The most contending economical hydrogen generation techniques utilizing visible light are driven from solar energy coupled with catalytic reactions to form photocatalytic processes. Though, this processes is promising, many techniques still suffers a challenge as best performing technique. It is quite less complex to split water \cite{takata2020photocatalytic, astruc2020introduction, chen2020engineering} into hydrogen (H\(_2\)) at cathode and oxygen (O\(_2\)) at anode from the coupling of solar (photons), and catalyst, known as the photocatalytic water splitting. 

Apart from sourcing photons from the abundance of lights from solar, near-infrared region, and light emitting diodes. These photocatalytic materials \cite{wu2023review, shen2020nanostructured, chen2020photocatalytic, shabdan2020photoactive, nuraje2015green} used for water splitting are predominately titanium dioxide (TiO\(_2\)), ferric oxide (Fe\(_2\)O\(_3\)), tungsten trioxide (WO\(_3\)), bismuth vanadate (BiVO\(_4\)) and zinc oxide (ZnO) among others. However, these materials are generally considered semiconductors with several different band gaps of which some could be heterogeneous, homogeneous or Plasmonic antenna-reactor structured type. TiO\(_2\) with a wide band gap absorbs photons from light spectrum of about 200 nm to 800 nm, this material excited electrons transition from its valence band to conduction band thereby creating holes in valence. The electrons and holes generated by the photo effect move towards the catalyst surface and actively engage in the process of reducing and oxidizing the water molecules that are adsorbed \cite{albero2020photocatalytic, wang2021advances, karimi2020tuning}. 

The fundamental engineering processes of photocatalytic quantum efficiency are mainly from light harvesting, surface adsorption capacity, charge separation, transport and utilization; Figure 1 highlights the different methods to photon-induced green hydrogen production. That said, it is reported by the United States Department of Energy (DOE) as of 2020 that solar to hydrogen (STH) \cite{zhou2023solar} conversion amounts to 20\% and has an ultimate target of 25\% improvement \cite{DOE2024}. Therefore, the continuous modeling of materials using the ab initio molecular dynamic principles and other predictive computational intelligence have been resorted to finding the best optimum structure for experimental photocatalytic processes to improving on light to hydrogen conversion (LTH).

The application of science and artificial intelligence (SciAI) \cite{xu2021artificial, zhang2021study} in material modeling for several hydrogen production methods such as direct air electrolysis (DAE), anion exchange membrane (AEM), proton exchange membrane (PEM), biomass gasification, microbial electrolysis cells, GenHydro reactor system, photo-electrochemical catalysis \cite{ursua2011hydrogen, ko2022direct} (PEC) to mention but a few have gained tremendous traction in the clean energy industry. This intelligent predictive and assessment of preferred heterogeneous catalysis have by far reduced the recent complexities and time associated to synthesizing materials in the laboratories. Large volumes of dataset for water splitting photocatalytic processes either from density function theory (DFT) calculations or from experimental analysis can be trained using machine learning or deep learning algorithms. While this is feasible, efforts are still at the elementary stage in the industry.

\begin{figure}[!ht]
    \centering
    \includegraphics[width=0.5\textwidth, angle=360]{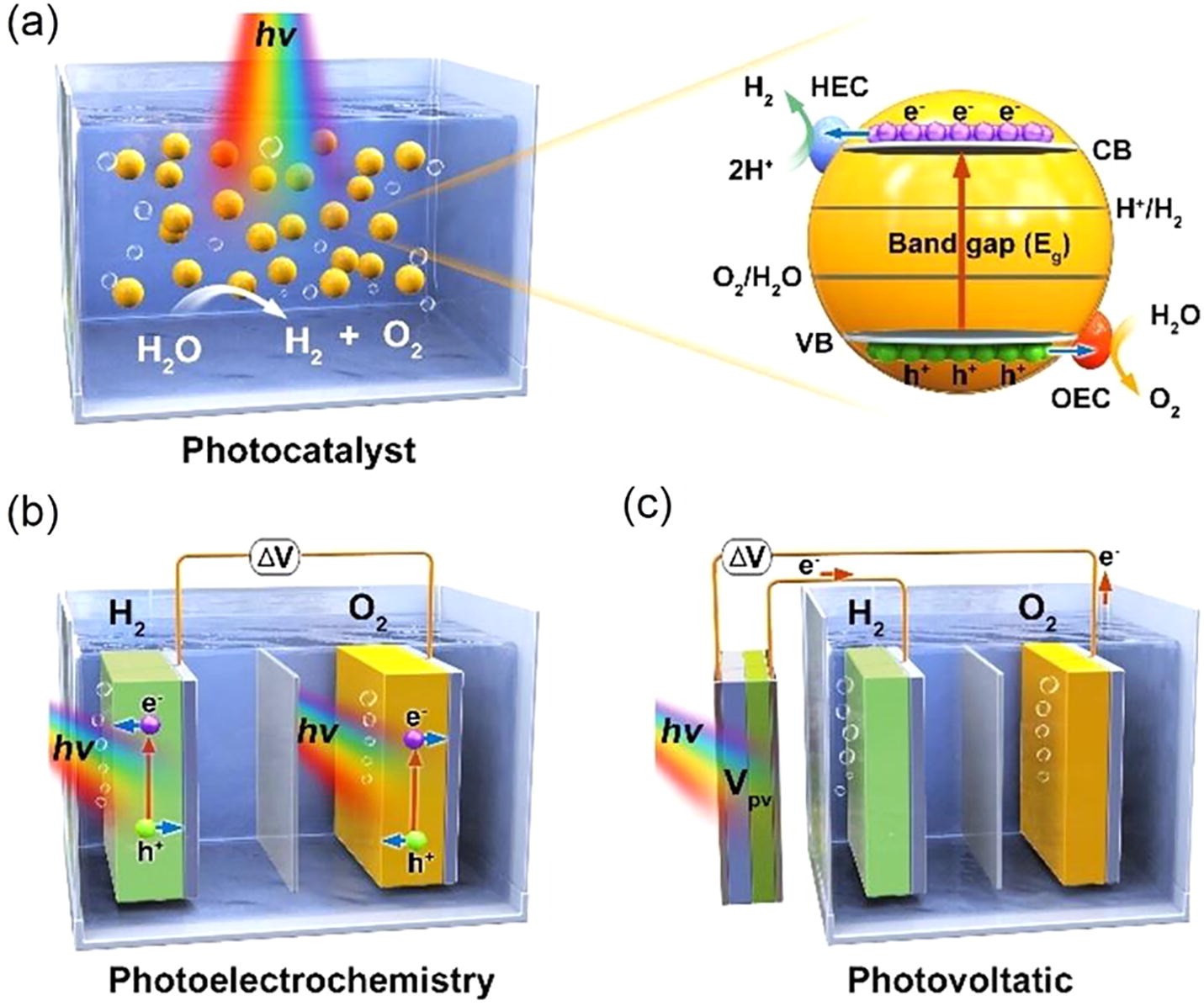}
    \caption{Illustration of three light-driven water splitting systems: (a) Photocatalytic (PC), (b) Photoelectrochemical (PEC), and (c) Photovoltaic-electrochemical (PV-EC). Each system demonstrates unique operational mechanisms, including photon absorption, charge separation, and redox reactions, offering insight into the diverse approaches for light-driven hydrogen generation, adapted with permission from Bian, 2021 \cite{bian2021perovskite}}
    \label{fig1}
\end{figure}

The significance of the input parameter is as important as tweaking hyperparameters of the algorithms to optimize process parameters. Since data plays a critical role for design optimization, ScienceDirect shows a total of 145,382 photocatalytic results \cite{ScienceDirect2024}. Sampling experimental or published data can include high degree of inconsistencies from different research groups with several process and measuring equipment, and this makes training and validation quite complex. For this reason, standardizing parameters for photocatalytic processes from any derived light source is a major discussion for the industry. That said, learning patterns of data with different architectures from regressions to convolutions \cite{kumar2021chemical, masood2019machine, isazawa2023automated} are explored in this study under several light to hydrogen (LTH) descriptors, such as energy band gap, absorption co-efficient, and electron energy loss function. 

Despite some data-driven inaccuracies for coupling artificial intelligence with material modeling efficiencies. This critical review takes a careful discussion on the evolution, recent development, challenges and perspectives of first principles molecular dynamics \cite{irawan2024molecular}, semiconductor materials and light sources, photocatalytic hydrogen production methods, machine and deep learning architectures. Figure 2 presents an informed framework and discussion to support photocatalytic experts of the recent intelligence to aid managerial decision making.

\begin{figure}[!ht]
    \centering
    \includegraphics[width=0.6\textwidth, angle=360]{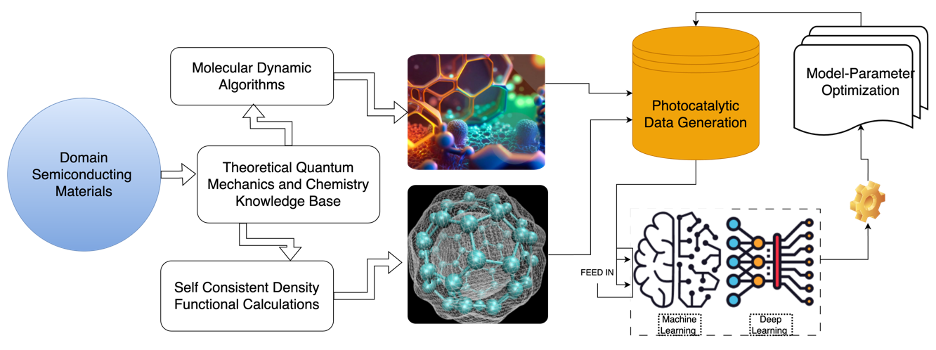}
    \caption{Feed Process of Photocatalytic Database from MD/DFT Codes to Machine and Deep Learning Algorithms}
    \label{fig2}
\end{figure}

%--------------------------------------------------------
\section{Ab Initio Molecular Dynamics (AIMD) and Density 
Functional Theory (DFT) Approach}
%--------------------------------------------------------

The experimental analysis of catalytics and their atomic interactions with lights are visible through spectroscopy. Backing up the theoretical processes with computational techniques is necessary to further examine the electronic structure such as phonon, energy
band gap, and electron energy loss. Despite the need to define the molecular dynamics \cite{isazawa2023automated, hollingsworth2018molecular} of photocatalysis from a computational viewpoint, several scientific programmers have introduced relevant algorithms to process intricate behaviour of electrons as a wave and particle. A quick search of molecular dynamics on GitHub shows 1,100 results \cite{leimkuhler2015molecular}. However finding the best algorithm to process the said elements could be challenging, since most of the deployed releases are not updated with the current widely accepted quantum mechanical approaches. On the contrary, some package releases are constantly updated to meet the trending needs from quantum chemistry and physics under accurate molecular dynamics (MD) and density functional theorem (DFT) algorithms.

Whereas ab initio MD focuses on the general fluctuation of atoms and molecules based on its natural state and origin, DFT pays attention to electron's ground state properties and its total energy from a fixed atomic position \cite{GitHub2024, orio2009density}. That said, DFT is now the most preferred theory for the investigation of semiconducting materials for photocatalytic processes. To begin with, it is highly recommended by research that before semiconducting materials are designed or modified into a photocatalytics, the structural, optical, magnetic and electronic properties  of the materials can be feasibly computed without having to go through sophisticated experimental analysis. This DFT model approach for simulating materials are based on quantum mechanics principles which does not heavily rely on experimental parameters.

Manipulating elements from the periodic table have led to several atomic exploitation to finding the best minimum energy configurations named as the ground state energy where the quantum number n=1. In further exploitation, most DFT codes focuses on is mostly on the self-consistency fields from Kohn-Sham under various widely accepted calculated forms of variable cell relaxations, relaxations, density of states and wave functions, charge density, band structure and phonon. However due to the inaccuracy of Coulombs potentials, pseudopotentials which is also known as the effective potentials are used to represent approximations for defining atomic complex systems. In these psuedopotentials approximations are kohn-sham exchange-correlation (XC) functionals \cite{Author82024}, which are known for defining the energy induced in the interactive electrons. The most popular and recent XC functionals remains the local density (LDA) \cite{kohn1965self}, generalized gradient (GGA) \cite{perdew1986density, perdew1992atoms, perdew1996generalized} and other hybrid functionals such as Perdew– Burke–Ernzerhof (PBE0) \cite{hammer1999improved}. These codes emerging from the  plane wave electronic calculation predominantly uses either ultrasoft or norm-conserving pseudopotentials. In most case, these computational codes requires higher central and graphic processing units CPU/GPU as this amounts to computational cost. As shown in Figure 3 Genovese et. al. \cite{genovese2009density} had experimented ways of reducing computational cost by running full electronic structure calculation code on a hybrid parallel architecture with Graphic Processing Units (GPU).

\begin{figure}[!ht]
    \centering
    \includegraphics[width=0.5\textwidth, angle=360]{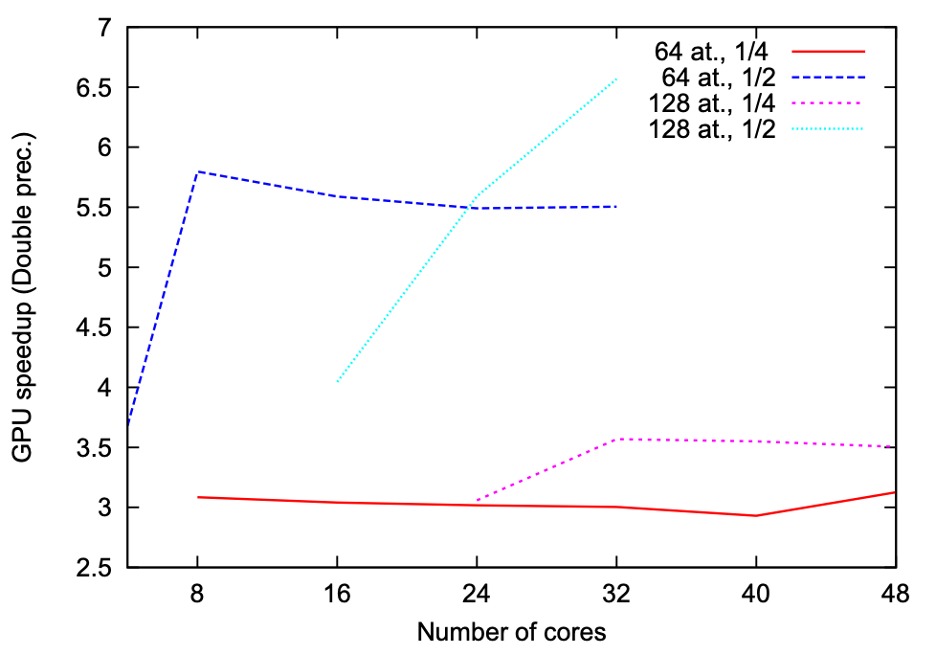}
    \caption{The integration of GPUs with CPU cores for running DFT codes. The plot highlights GPU speedup trends for varying atom sets, emphasizing computational efficiency improvements when optimizing GPU-CPU configurations, adapted with permission from Genovese, 2009 \cite{genovese2009density}}
    \label{fig3}
\end{figure}

The said experiment was run from BIGDFT codes with emphasis on the overall communication between number of CPU cores with graphic card (many cores to one card). Similar GPU speedup is observed in red line throughout 68 atoms 1/4 repartition despite the number of CPU cores. In similar faction, the GPU speedup for 128 atoms 1/4 repartition remained same after 32 cores. Optimizing the number of cores to one graphic card may be dependent with different algorithms, but this demonstration foretells that a single graphic card and eight CPU cores is quite enough for calculating complex full DFT codes.

The importance of modelling materials for their photocatalytic capabilities is to define their band gap. The larger the band gap, the better the semiconducting material suitable for catalysis; on the contrary, a smaller band gap indicates conductivity. Since this is the most crucial hyperparameter for modelling semiconducting materials, it is most often underestimated in DFT calculations \cite{perdew1985density}. The reason is that band gaps from DFT codes do not correlate with experimental results. On many occasions, this has been attributed to errors in electronic band structure codes that do not satisfy the true empirical solution. However, when determining the overall energy of the Nth electrons in a system, as proposed by Sham et al. \cite{sham1983density}, it basically involves comparing these energy levels of the ground state. Consequently, the variation in energy between the highest point in the valence band and the lowest point in the conduction band is referred to as the band gap. Despite the many attempts to validate this with experimental solutions, modified hybrid functionals such as Heyd-Scuseria-Ernz-Erhof HSE03/HSE06 and PBE0 are seen to correct the DFT codes for obtaining appropriate matching empirical band gap solutions. Whereas PBE0 maintains an exact Hartree–Fock (HF) exchange, HSE03 and HSE06 consider the expansion of HF calculations. It is worth mentioning that LDA and GGA are deemed unreliable for calculations, whereas HSE and PBE modified functionals are considered the most dependable. Though this raises the cost of computation and simulation time, the modified hybdridal functionals are popularly used.

Compiling algorithms and running them in different environments with graphical user interfaces throws wild scientific programming skills at developers. With the advances in quantum mechanics and chemistry, there are currently about 54 different material modeling software programs developed. About 90\% of these run on the FORTRAN compiler, an imperative scientific programming language, while the others run on Python, C, C++, CUDA, and Perl. However, running this software mostly requires some level of programming skills, so release and install the appropriate packages from repositories through the computers' terminal. This challenges a lot more students, and few fall for the simulation interest. Whereas commercially licensed software provides intuitive graphical interactions with the software, most free software under LGPL, GPL, MIT, ECL, and BSD licences will have to be run in the most difficult way for non-programmers. Moreover, out of the 54 software programs, 26 are considered to have the freedom to run, study, share, and modify the software, except for the MIT license, which has fewer exceptions, though it is free. Narrowing down the free simulation software to molecular dynamics (MD), density functional theory (DFT), and time-dependent density functional theory (TDDFT) calculations, BigDFT \cite{genoveseimprovements}, ABINIT \cite{gonze2009abinit}, Quantum ESPRESSO \cite{giannozzi2009quantum}, and SIESTA \cite{soler2002siesta} are the most suitable ones for simulations. It is worth noting that the free version of SIESTA ends in 2003, subsequent updated versions are for academic and commercial purposes only. Establishing free software for complex material modeling brings relief to the lower income group of scientists, enabling them to expand their knowledge to use and contribute to the software development. That said, out of these 54 software licenses, 16 are academic licenses. In a similar faction, the following are able to perform all three verifiable computational combinations: MD, DFT, and TDDFT; VASP \cite{hafner2008ab}, ASAP-SIESTA \cite{ASAPSIESTA47}, CASTEP \cite{clark2005first}, Q-Chem \cite{shao2015advances}, and ORCA \cite{neese2012orca}.

The most popular academic software and widely used material modeling software are the Spanish Initiative for Electronic Simulations with Thousands of Atoms (SIESTA), the Vienna Ab initio Simulation Package (VASP), and the Cambridge Serial Total Energy Package (CASTEP). Codes from this software generally focus on catalysis, solids and molecules, phase and structure stability, quasi-crystals, magnetic nanostructures, semiconductors, insulators, interfaces, surfaces, thin films, and chemical processes.

SIESTA, known for its robust computational material modeling algorithms, uses the FORTRAN 95 compiler for electronic structure and molecular dynamics calculations. It also presents self-consistent field calculations from Kohn-Sham's DFT calculations. With over 10,000 atoms, SIESTA computation surpasses the codes from plane-wave algorithms. Atomistic Simulation Advanced Platform (ASAP) is presently responsible for running the SIESTA solver, which provides a ready-to-use package for non-programmers and is suitable for all MD, DFT, and TDDFT calculations for solids and molecules. Some of their workflows include single point, geometry optimization, molecular dynamics, equation of state, automated convergence tools (BZ sampling, mesh cutoff), nudged elastic band, phonons and vibrations, optical response, and electronic transport calculations.

CASTEP is a cutting-edge quantum mechanics-based programme created especially for fields where empirical models and experimental data may be sparse, such as solid-state physics, chemical engineering, and materials research. In order to investigate the characteristics of surfaces and crystals in materials such metals, semiconductors, ceramics, and zeolite, CASTEP uses the DFT plane-wave pseudopotential approach.

VASP is a computer programme that uses a plane wave basis set and either the PAW method or Vanderbilt pseudopotentials to explain the electron–core interaction for atomic scale materials modelling for quantum-mechanical molecular dynamics and electronic structure calculations. Using DFT to solve the Kohn-Sham equations or the HF approximation to solve the Roothaan equations, VASP approximates the solution to the many-body Schrödinger problem. For solid-state, molecular, polymer, and surface applications, the performance of VASP is comparable to Gaussian-type DFT implementations. The VASP software employed various post-DFT adjustments, such as HF exchange, many-body perturbation theory, hybrid functionals, and dynamic electronic correlations within the random phase approximation. These enhancements were utilized alongside the core DFT method.

Many of the software discussed can run on local machines, ASAP-SIESTA  is known for both local and remote calculations, whereas Mat3ra \cite{mat3ra51} provides an interactive cloud-based platform for material modelling. Figure 4 presents a schematic view of the software discussed.  

\begin{figure}[!ht]
    \centering
    \includegraphics[width=0.5\textwidth, angle=360]{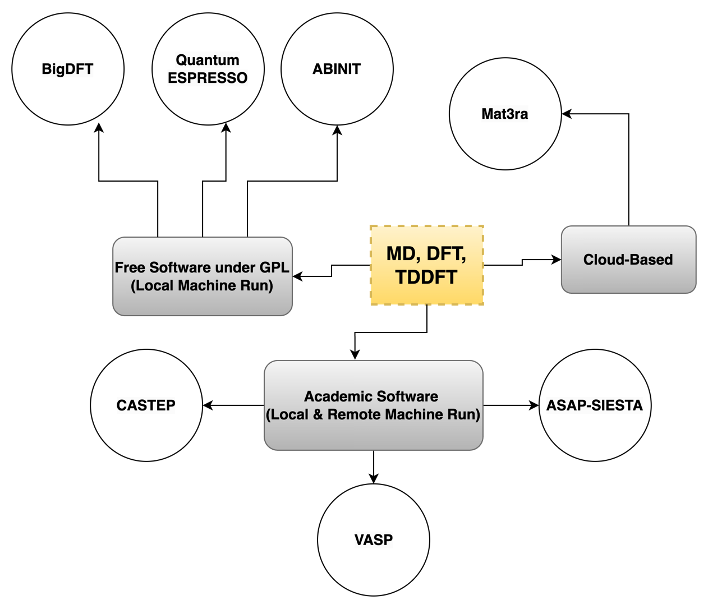}
    \caption{Schematic flow of recent free and academic software licenses for molecular dynamics, density functional theory and time-dependent density functional theory calculations}
    \label{fig4}
\end{figure}

%--------------------------------------------------------
\section{Evolution of Light-driven Photoreactions}
%--------------------------------------------------------

Light-driven photoreactions are the cornerstone of photocatalytic water splitting, offering sustainable avenues to harness solar energy for hydrogen production. Over the years, the progression of light sources and materials has significantly shaped the field, particularly through advances in photon absorption efficiency and energy conversion pathways. This section delves into the evolution of light sources—ranging from solar and near-infrared (NIR) radiation to mercury discharge lamps and light-emitting diodes (LEDs)—and their applications in photocatalysis.

%--------------------------------------------------------
\subsection{Solar-NIR Radiation}
%--------------------------------------------------------
The utilization of solar energy in the photocatalytic splitting of water has been pivotal in advancing clean energy technologies. Solar energy is abundant, cost-free, and inherently sustainable, making it the most attractive light source for large-scale hydrogen production. Traditional photocatalytic materials, such as titanium dioxide (TiO\(_2\)), were initially developed to absorb ultraviolet (UV) light, which comprises only about 4–5\% of the solar spectrum. However, as research advanced, it became evident that harnessing the visible and near-infrared (NIR) portions of the spectrum was critical for improving overall solar-to-hydrogen (STH) efficiency \cite{kumar2024production, agrawal2024green, priya2024photocatalytic, patel2024hydrogen}. Modern solar simulators are of late considered a practical step to simulating electrochemical cells for water splitting in the laboratory; Figure 5 showcases a typical solar-simulator in the University College London for water splitting. 

\begin{figure}[!ht]
    \centering
    \includegraphics[width=0.6\textwidth, angle=360]{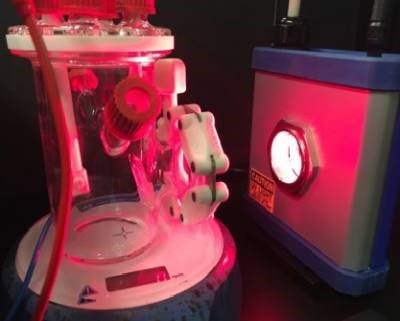}
    \caption{Solar simulator-driven photoelectrochemical water splitting, adapted from web \cite{ucl2024}}
    \label{fig5}
\end{figure}
 
Recent innovations in material science \cite{ansari2021advancement, cao2022recent} have enabled the development of semiconductors and doped materials capable of extending light absorption into the NIR range (700–2500 nm). Upconversion nanoparticles (UCNPs), for instance, have emerged as effective mediators that convert NIR photons into higher-energy visible photons. By integrating UCNPs with conventional photocatalysts such as TiO\(_2\) or WO\(_3\), researchers have achieved significant improvements in water-splitting efficiency under NIR irradiation.

Moreover, tandem systems employing multi-junction photocatalytic materials have demonstrated improved photon utilization by sequentially absorbing photons across the UV, visible, and NIR regions. For example, bismuth vanadate (BiVO\(_4\)) coupled with a silicon-based photoelectrode has shown promise in maximizing STH conversion \cite{jin2022two}.

Beyond photon absorption, the photothermal effects \cite{qureshi2024advances} of NIR radiation have gained attention. Photothermal catalysis leverages localized heating effects induced by NIR photons to enhance reaction kinetics at the catalyst surface. Plasmonic nanomaterials, such as gold and silver nanoparticles, are often used to amplify these effects, leading to improved hydrogen evolution rates.

%--------------------------------------------------------
\subsection{Mercury Discharge Lamps}
%--------------------------------------------------------

Historically, mercury discharge lamps played a pivotal role in the early stages of photocatalytic research. These lamps emit a broad spectrum of light, including UV, which was instrumental in validating the photocatalytic activity of materials like TiO\(_2\).

Mercury discharge lamps offer high UV intensity \cite{broxtermann2017mercury}, making them suitable for studying UV-responsive photocatalysts. However, their inefficiency in energy consumption and environmental concerns related to mercury disposal have limited their modern-day applicability. Despite this, these lamps remain valuable tools as elucidated in Figure 6 for laboratory-scale experiments that require precise control over UV-vis light exposure \cite{kubiak2024impact, meinhardova2023role, horikoshi2019development}.

\begin{figure}[!ht]
    \centering
    \includegraphics[width=0.6\textwidth, angle=360]{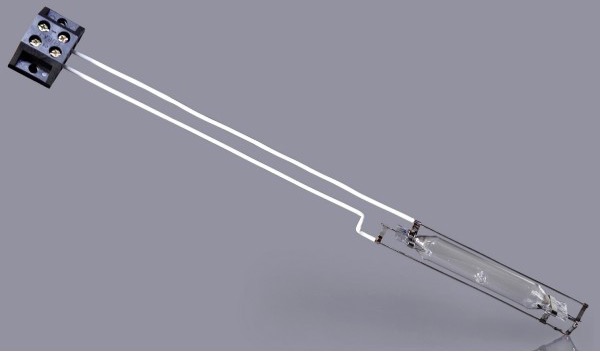}
    \caption{Image of mercury lamps by industries with claims to be versatile in emitting lights at an excited state with visible spectrum ranging from 200nm to 800nm, adapted from web \cite{shilpent2024}}
    \label{fig6}
\end{figure}

One notable application of mercury discharge lamps is their use in evaluating quantum efficiencies under controlled conditions. Researchers  have used these lamps to identify reaction mechanisms by isolating the effects of UV-induced charge carriers. However, the lack of compatibility with visible or NIR-responsive photocatalysts has led to a gradual decline in their use, particularly as the field transitions to more sustainable and efficient light sources.

%--------------------------------------------------------
\subsection{Light Emitting Diodes (LEDs)}
%--------------------------------------------------------

The advent of light-emitting diodes (LEDs) \cite{bhattarai2024review} has revolutionized photocatalysis by providing an energy-efficient, tunable, and environmentally friendly alternative to traditional light sources. LEDs offer narrow-band emission, enabling researchers to tailor light wavelengths to match the absorption spectra of specific photocatalysts.

LEDs consume significantly less power than mercury discharge lamps and have a longer operational lifespan. Additionally, their ability to emit light across the UV, visible, and NIR spectra makes them versatile for a wide range of photocatalytic applications. By utilizing LEDs with customizable wavelengths, researchers can optimize photocatalytic processes for various materials and reaction conditions. For example, blue and red LEDs are often paired with visible-light-responsive photocatalysts like BiVO\(_4\) and cadmium sulfide (CdS), respectively.

Recent studies have demonstrated the integration of LED arrays in scalable photocatalytic reactors as demonstrated in Figure 7. Such systems combine the precision of wavelength control with the ability to uniformly illuminate large reaction surfaces. Additionally, the pulsed operation of LEDs has been explored as a method to enhance charge separation in photocatalysts. Pulsed LED systems intermittently provide light to the reaction system, allowing time for photogenerated charge carriers to participate in redox reactions before recombining.

\begin{figure}[!ht]
    \centering
    \includegraphics[width=0.5\textwidth, angle=360]{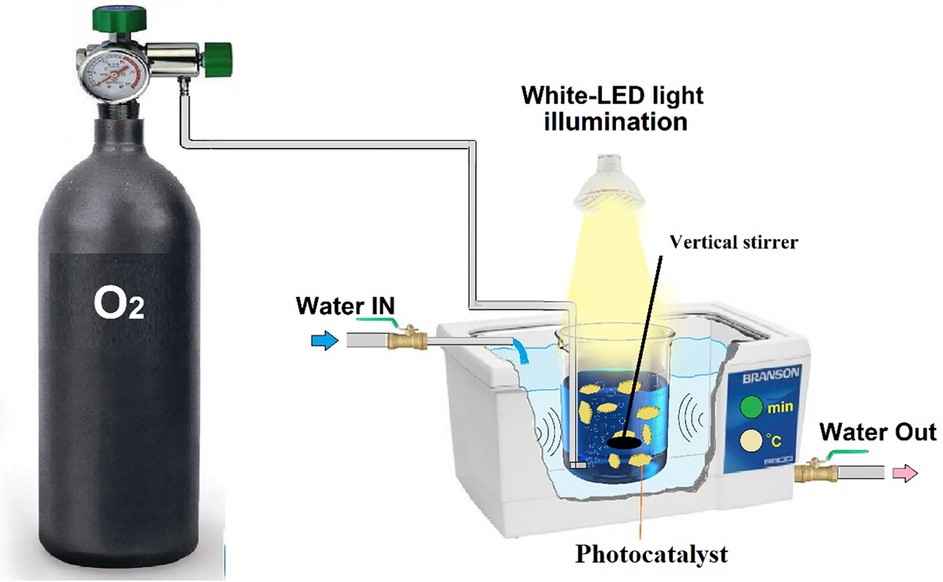}
    \caption{Water splitting process flow under white-LEDs and storage, adapted with permission from Sareshkeh, 2023 \cite{tarighati2023preparation}}
    \label{fig7}
\end{figure} %open access article

The evolution of light-driven photoreactions is closely tied to the development of photocatalytic materials optimized for specific light sources \cite{khodadadian2019optimizing, balarabe2022visible}. For instance, UV-responsive materials like TiO\(_2\) were initially the primary focus, but advancements in doping, heterojunction formation, and plasmonic effects have expanded the applicability of materials to the visible and NIR regions. Furthermore, hybrid light sources that combine solar simulators with LEDs have been employed to mimic real-world conditions, enabling more accurate assessments of photocatalytic performance.

%--------------------------------------------------------
\subsubsection{Emerging Trends in Light Sources}
%--------------------------------------------------------
Perovskite materials \cite{lim2019perovskite, zhu2019lead, gadgil2016photocatalytic}, known for their tunable optical properties, are emerging as novel light sources for photocatalysis. Their high quantum yield and spectral tunability make them attractive for coupling with semiconducting photocatalysts. Wireless light systems, such as laser-induced photocatalysis, are also gaining traction. These systems use high-intensity lasers to activate specific regions of a catalyst, enabling localized reaction control. While still in the experimental stage, this approach offers exciting possibilities for precision-driven photocatalysis.

%--------------------------------------------------------
\section{Photocatalytic Semiconductor Materials}
%--------------------------------------------------------
Semiconductors form the backbone of photocatalytic water splitting \cite{markhabayeva2019designing}, with their ability to absorb photons and generate charge carriers (electrons and holes) crucial to the hydrogen evolution reaction (HER) and oxygen evolution reaction (OER). Over the years, significant advancements have been made in the synthesis, doping, and modification of these materials to improve their efficiency, stability, and adaptability to various light sources. This section explores the characteristics, recent innovations, and challenges associated with key semiconductor materials used in photocatalysis \cite{goodarzi2023recent}.

Titanium dioxide is the most widely studied and utilized photocatalyst, owing to its stability, abundance, and non-toxicity. Its bandgap of ~3.2 eV allows it to absorb ultraviolet (UV) light, which limits its utilization of the solar spectrum but makes it effective for controlled experiments. To overcome the limitations of TiO\(_2\)’s UV specificity, researchers have focused on doping and surface modifications. Metal doping (e.g., Ag, Fe) introduces intermediate energy states within the bandgap, enabling visible light absorption. Non-metal doping (e.g., N, S) shifts the bandgap edge by modifying the electronic structure, improving light absorption in the visible spectrum. Heterojunction formation, combining TiO\(_2\) with other semiconductors such as WO\(_3\) or ZnO, enhances charge separation and reduces recombination rates. Nanoscale TiO\(_2\) (e.g., nanotubes, nanowires, and nanoparticles) has been extensively studied due to its high surface area and enhanced photoreactivity. The synthesis of vertically aligned TiO\(_2\) nanotube arrays, for instance, has shown promise in improving electron transport efficiency \cite{yang2024modified, miyoshi2018water}.

WO\(_3\) \cite{wang2012rational}, with a bandgap of 2.6–2.8 eV, is a visible-light-responsive semiconductor known for its stability under acidic conditions. WO\(_3\)’s ability to drive oxygen evolution reactions efficiently has made it an essential component in tandem systems. Coupling WO\(_3\) with hydrogen-evolution catalysts \cite{sarkar2024situ} such as Pt improves overall water-splitting efficiency. Plasmonic nanoparticles (e.g., Au, Ag) integrated with WO\(_3\) create localized surface plasmon resonance (LSPR), boosting visible light absorption and promoting hot-electron transfer.

ZnO, similar to TiO\(_2\), is a UV-active semiconductor with a bandgap of ~3.37 eV. While it shares many properties with TiO\(_2\), ZnO is more susceptible to photocorrosion, limiting its long-term applicability. Researchers have developed composite systems, combining ZnO with carbon-based materials (e.g., graphene, carbon nanotubes) to mitigate photocorrosion and enhance charge transport. ZnO is often paired with other semiconductors in binary systems to improve stability and broaden light absorption \cite{abou2024advancements}. For example, ZnO/CdS heterojunctions have demonstrated enhanced photocatalytic performance under visible light \cite{qi2023enhanced}.

BiVO\(_4\) has gained prominence for its ability to absorb visible light (bandgap ~2.4 eV) and its compatibility with water oxidation reactions \cite{tolod2020optimization}. Monoclinic BiVO\(_4\) exhibits superior photoactivity \cite{nuraje2021photoactive} compared to tetragonal or scheelite structures. Modifications such as introducing oxygen vacancies or forming heterojunctions with materials like g-C\(_3\)N\(_4\) have further improved its catalytic properties. BiVO\(_4\) is frequently used in tandem PEC cells, where it serves as a photoanode, complementing hydrogen-evolution photocathodes. This configuration maximizes solar-to-hydrogen conversion efficiency.

%--------------------------------------------------------
\subsection{Emerging Binary Semiconductor Materials} 
%--------------------------------------------------------
Binary semiconductor systems are gaining traction due to their synergistic properties. These materials combine the strengths of individual components, such as enhanced light absorption, improved charge separation, and stability \cite{meyer2012binary}.

The combination of cadmium sulfide (CdS) and zinc sulfide (ZnS) in core-shell structures helps reduce photocorrosion and improve charge separation as demonstrated by Vamvasakis et. al. \cite{vamvasakis2023mesoporous} in Figure 8. This system is particularly effective under visible light, achieving high hydrogen evolution rates \cite{jiang2016cocatalyst}. As a metal-free semiconductor, graphitic carbon nitride (g-C\(_3\)N\(_4\)) offers a unique approach to photocatalysis as in the case of Wang et. al. \cite{wang2024two}. Its bandgap of around 2.7 eV allows it to absorb visible light, and its compatibility with various co-catalysts makes it versatile. Coupling g-C\(_3\)N\(_4\) with noble metals like platinum or ruthenium significantly enhances its photocatalytic activity \cite{vasilchenko2022platinum}. Perovskite materials, with their adjustable bandgaps and excellent light absorption properties, are emerging as promising candidates for high-efficiency photocatalysis. However, stability issues continue to be a significant challenge.

\begin{figure}[!ht]
    \centering
    \includegraphics[width=0.6\textwidth, angle=360]{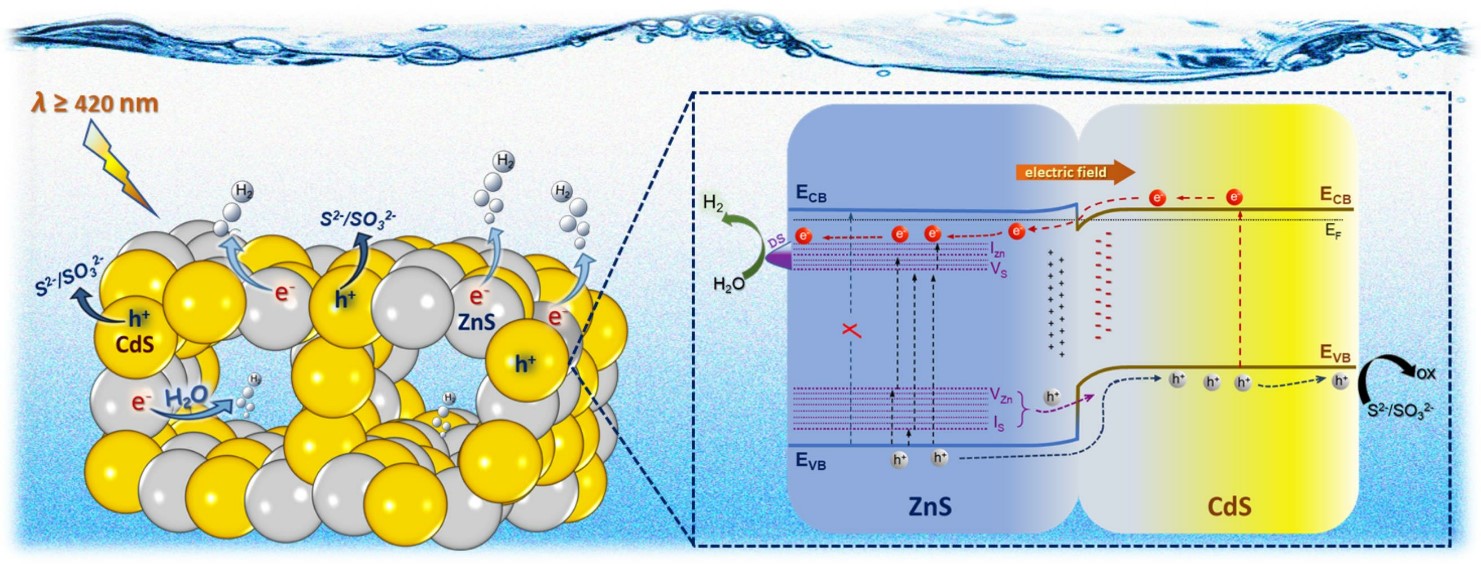}
    \caption{Demonstration of CdS/ZnS as a binary semiconducting material under visible light, adapted with permission from Vamvasakis, 2023 \cite{vamvasakis2023mesoporous}}
    \label{fig8}
\end{figure} %open access article

Developing semiconductor materials for photocatalysis still faces several challenges, despite progress. Achieving the right balance between light absorption and charge carrier movement in the bandgap remains a key issue. Many semiconductors, especially those sensitive to visible light, are prone to photodamage or deterioration during reactions. The rapid recombination of the generated electrons and holes reduces overall efficiency, requiring innovative approaches to separate charges. Scaling up the production of high-performing semiconductors while maintaining consistency and cost-effectiveness is a significant obstacle \cite{villa2021photocatalytic}.

Research is now concentrating on combining improved materials and methods to solve these problems. Using machine learning in material design allows for predictive modeling and optimization, which is speeding up the discovery of new semiconductor materials. The use of quantum dots as light-absorbing agents shows potential for expanding light absorption into the near-infrared region. Combining organic and inorganic materials in hybrid systems provides a way to achieve both high efficiency and stability in photocatalysts \cite{ma2023perspective}.

%--------------------------------------------------------
\subsection{Hybrid Quantum Dots and Upconversion Materials} 
%--------------------------------------------------------
Hybrid quantum dots (QDs) and upconversion materials \cite{gui2024application, chen2022up} represent cutting-edge advancements in the field of photocatalytic semiconductors. These materials are designed to overcome the limitations of conventional semiconductors, such as restricted light absorption and inefficient charge carrier dynamics. By combining quantum dots with upconversion nanoparticles, researchers have created hybrid systems capable of extending light absorption into the near-infrared (NIR) region and enhancing photocatalytic efficiency.

%--------------------------------------------------------
\subsubsection{Quantum Dots in Photocatalysis}
%--------------------------------------------------------
Quantum dots \cite{jacak2013quantum} are tiny semiconductor particles that exhibit unique size-dependent optical and electronic properties. Their adjustable bandgaps allow for precise control over light absorption, making them highly effective for photocatalytic applications. These dots can absorb light across a wide range of the spectrum, including visible and near-infrared regions. Their small size reduces charge recombination rates, improving overall photocatalytic performance. Surface modifications with ligands or co-catalysts can enhance their interaction with water molecules and reaction intermediates. While cadmium-based quantum dots, such as CdS and CdSe, have been extensively studied for their high quantum efficiency \cite{gomes2011cdte}, concerns over toxicity have led to the development of alternative, lead-free \cite{leng2016lead} options like ZnS and CuInS\(_2)\), which are more environmentally friendly.

%--------------------------------------------------------
\subsubsection{Upconversion Materials}
%--------------------------------------------------------
Upconversion materials can convert low-energy near-infrared (NIR) light into higher-energy visible light, allowing the activation of photocatalysts that cannot be activated by NIR light alone. These materials typically consist of rare-earth-doped nanoparticles, such as NaYF\(_4\) doped with ytterbium (Yb\(^{3+}\)) and erbium (Er\(^{3+}\)) ions. The energy transfer upconversion (ETU) process involves the absorption of multiple low-energy photons, which leads to the emission of a single high-energy photon. The photon avalanche is a cascade of energy transfer events that amplifies the upconversion process. Upconversion nanoparticles are combined with semiconductors like TiO\(_2\), BiVO\(_4\), or WO\(_3\) to expand their activity into the NIR region. This hybrid system improves the utilization of the solar spectrum and enhances photocatalytic efficiency \cite{zhou2015upconversion}.

%--------------------------------------------------------
\subsubsection{Hybrid Quantum Dot-Upconversion Systems}
%--------------------------------------------------------
Combining quantum dots and upconversion materials creates a synergistic system that leverages the strengths of both components. Upconversion nanoparticles act as a foundation, converting near-infrared light to visible light, which is then absorbed by the quantum dots. Alternatively, quantum dots and upconversion materials are integrated into a single nanostructure, enabling efficient energy transfer and reducing recombination losses. These hybrid systems effectively enhance light-driven water splitting reactions. Quantum dots capture high-energy photons for electron generation, while upconversion materials provide additional excitation under near-infrared light, improving the hydrogen evolution reaction. Furthermore, the improved charge separation and reduced recombination significantly boost the efficiency of the oxygen evolution reaction.

Hybrid QD-upconversion systems have great potential, but there are still challenges to overcome. The use of cadmium and lead in quantum dots raises environmental concerns, so non-toxic alternatives must be developed. Improving the efficiency of energy transfer between upconversion materials and quantum dots is crucial to minimize losses. Scaling up the production of these hybrid systems while maintaining consistency and performance is a significant obstacle \cite{liu2018synergistic, doughan2015paper}. The solution lies in developing non-toxic, widely available materials, exploring advanced fabrication methods like self-assembly to enhance scalability, and integrating machine learning to optimize the design and performance of hybrid systems.

%--------------------------------------------------------
\section{Water Splitting Mechanisms}
%--------------------------------------------------------
Photocatalytic water splitting represents a crucial process for sustainable hydrogen production, involving the decomposition of water molecules into hydrogen (H\(_2\)) and oxygen (O\(_2\)) under light irradiation. This section explores the key mechanisms governing water splitting, focusing on photon absorption, charge separation and transport, and surface reactions. Special attention is given to the roles of photocatalysts, co-catalysts, and reaction interfaces in enhancing efficiency and addressing challenges such as charge recombination and catalyst degradation. 

Photocatalysts absorb incident photons with energy equal to or greater than their bandgap, promoting electrons from the valence band (VB) to the conduction band (CB). This creates electron-hole pairs that drive the subsequent redox reactions. Materials with a suitable bandgap (1.8–3.0 eV) can efficiently utilize sunlight. Wide-bandgap semiconductors like TiO\(_2\) primarily absorb UV light, while doped and heterostructured materials \cite{moniruddin2016hierarchical, afroz2018heterojunction} extend absorption into the visible and NIR regions.

Efficient charge separation is essential for preventing recombination and ensuring the availability of electrons and holes for redox reactions. Excited electrons in the CB migrate to the photocatalyst surface to participate in the hydrogen evolution reaction (HER). Holes in the VB oxidize water molecules, driving the oxygen evolution reaction (OER). The efficiency of this process is determined by the lifetime of the charge carriers and the conductivity of the photocatalyst. Nanostructuring, co-catalyst integration, and heterojunction formation are common strategies to enhance charge separation and transport. Water molecules adsorbed on the photocatalyst surface undergo reduction and oxidation reactions \cite{odabasi2023strategies}:

\begin{equation}
    HER: 2H^+ + 2e^- \rightarrow H_2
\end{equation}

\begin{equation}
    OER: 2H_2O + 4h^+ \rightarrow O_2 + 4H^+
\end{equation}

The rates of these reactions depend on the surface area, active site availability, and catalytic efficiency of the material.

%--------------------------------------------------------
\subsection{Role of Co-Catalysts} 
%--------------------------------------------------------
Co-catalysts play a pivotal role in enhancing the efficiency of photocatalytic water splitting by facilitating charge separation and catalyzing surface reactions \cite{tian2023cocatalysts}. Noble metals like platinum (Pt) are highly effective hydrogen evolution co-Catalysts (HER) co-catalysts due to their low overpotential and high conductivity. However, cost and scarcity have driven the exploration of alternatives such as transition metal phosphides (Ni\(_2\)P) and Sulfides (MoS\(_2\)). Oxygen evolution co-catalysts (OER) is often the rate-limiting step in water splitting. Co-catalysts such as IrO\(_2\) and RuO\(_2\) are commonly used but are expensive. Emerging options include cobalt-based catalysts (Co\(_3\)O\(_4\)) and Perovskite Oxides such as LaNiO\(_3\). Qi et al. illustrates in Figure 9 how a two-step photoexcitation under a co-catalyst occurs with an aqueous redox mediator enhances water splitting processes, whereas Zhang et al. demonstrates energy diagram for PEC water splitting reaction in Figure 10 using an n-type semiconductor photoanode and a counter electrode immersed in an electrolyte.

%--------------------------------------------------------
\subsection{Heterojunctions in Charge Separation} 
%--------------------------------------------------------
Heterojunctions \cite{wang2024two} are interfaces between different semiconductors, designed to separate charges by creating an internal electric field. One common configuration is the Type-II Heterojunction, which aligns the conduction and valence bands of two semiconductors to enhance charge separation, such as TiO\(_2\) coupled with CdS. Another type is the Z-Scheme System, which mimics natural photosynthesis by using two semiconductors with complementary band structures, where electrons and holes from the two photocatalysts recombine, leaving highly energetic charge carriers for redox reactions, as seen in systems combining BiVO\(_4\) and g-C\(_3\)N\(_4\), which exhibit superior charge separation and redox capabilities.

\begin{figure}[!ht]
    \centering
    \includegraphics[width=0.5\textwidth, angle=360]{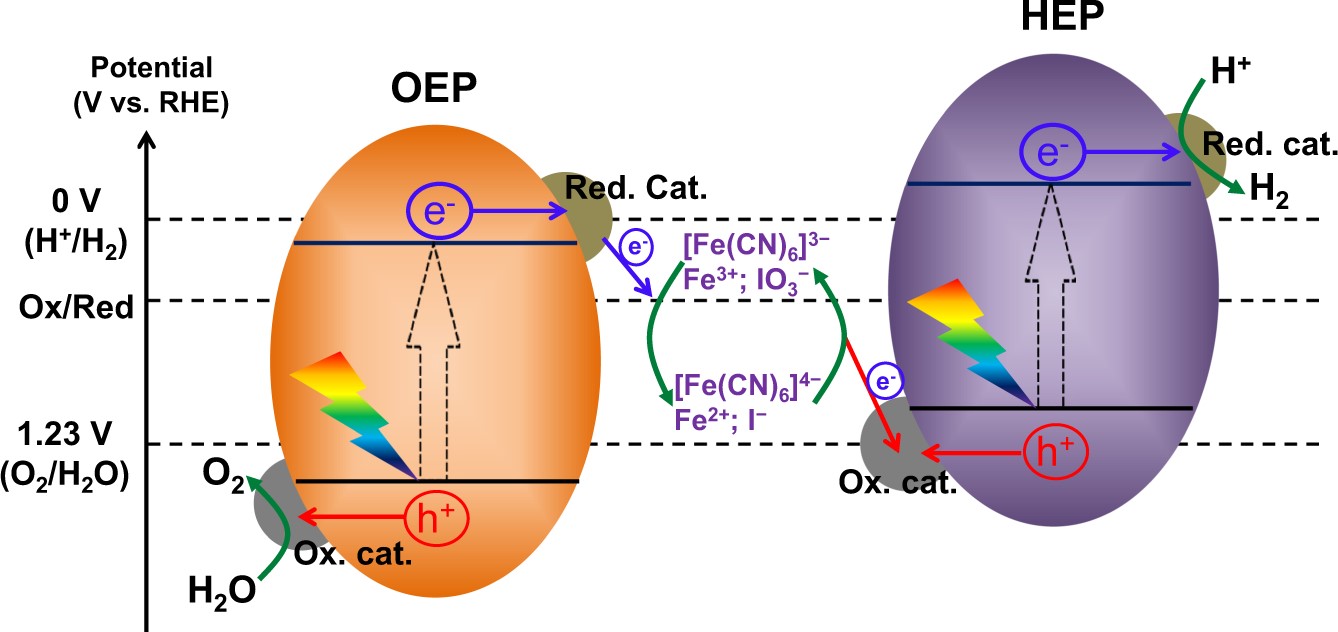}
    \caption{Illustration of Z-scheme energy diagram co-catalyst, adapted with permission from Qi, 2022 \cite{qi2022unraveling}}
    \label{fig9}
\end{figure} %open access article

Despite significant advancements, several challenges remain in achieving efficient and scalable water splitting. These include charge recombination, which reduces efficiency, and can be addressed by using co-catalysts to capture charge carriers and designing nanostructures to shorten charge carrier pathways. Catalyst stability is another challenge, as photocatalysts often degrade due to photocorrosion or oxidation, but can be addressed with protective coatings and stable co-catalysts. Additionally, both the hydrogen evolution reaction (HER) and oxygen evolution reaction (OER) require high overpotentials, increasing energy consumption, so developing catalysts with lower overpotentials is crucial. Finally, optimizing surface properties to enhance water molecule adsorption and reaction rates is a key area of focus.

\begin{figure}[!ht]
    \centering
    \includegraphics[width=0.4\textwidth, angle=360]{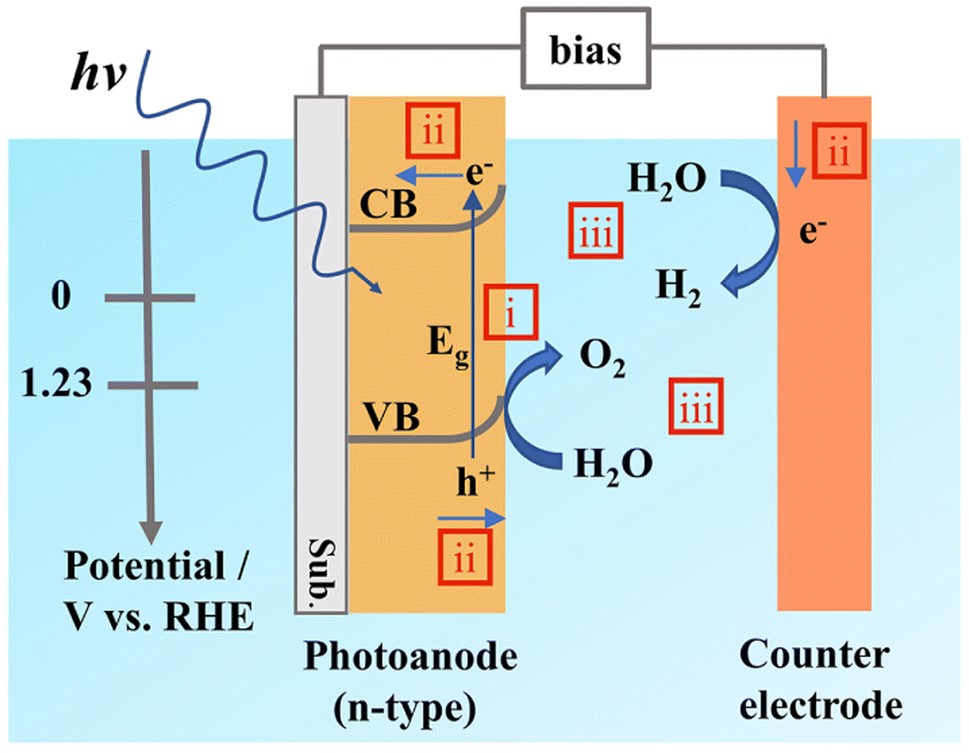}
    \caption{Photoelectrochemical (PEC) water splitting involves three steps: (i) light excites electrons, (ii) charges move through the material, and (iii) electrons are injected at the surface for the reaction, adapted with permission from Zhang, 2024 \cite{zhang2024effective}}
    \label{fig10}
\end{figure} %open access article

%--------------------------------------------------------
\subsection{Advanced Mechanisms and Emerging Trends} 
%--------------------------------------------------------
Plasmonic nanoparticles, such as gold (Au) and silver (Ag) \cite{lee2006gold, link1999alloy}, enhance light absorption through localized surface plasmon resonance (LSPR) \cite{sarina2013photocatalysis}. This generates hot electrons that participate in redox reactions. Plasmonic photocatalysis is particularly promising for NIR-responsive systems \cite{lee2024high}. Catalysts capable of performing both HER and OER on the same surface are being developed to simplify reaction systems and reduce costs. For instance, MoS\(_2\) modified with Ni(OH)\(_2\)) has shown dual functionality \cite{subramanian2018plasmon}. Photothermal catalysis utilizes localized heating effects to accelerate reaction kinetics. This approach is often combined with plasmonic systems to enhance efficiency.

To address current issues and make the technology commercially viable, research should concentrate on incorporating AI and machine learning to use predictive algorithms for finding ideal catalyst designs and reaction conditions, exploring abundant materials and scalable production methods to develop cost-effective catalysts, and designing multi-photon systems that can enhance light absorption.

%--------------------------------------------------------
\section{Recent Machine Learning Algorithms}
%--------------------------------------------------------
Machine learning (ML) as applied to many other fields \cite{campos2024evolutionary, wayo2023data, wayo2024filter} has emerged as a transformative tool in photocatalytic research, enabling the accelerated discovery, optimization, and deployment of efficient materials and processes with four staged essential paradigms as illustrated by Li et al in Figure 11 for machine learning. ML algorithms leverage large datasets from experiments and simulations to identify trends, predict outcomes, and optimize parameters, thereby reducing reliance on trial-and-error approaches. This section explores the latest advancements in ML applications for photocatalytic water splitting, focusing on regression, classification, and generative models, as well as reinforcement learning techniques.

Machine learning bridges the gap between material science and data-driven optimization by identifying correlations in complex datasets. Key contributions of ML in photocatalysis include predicting bandgaps, charge mobility, and stability of potential photocatalysts using computational models such as Density Functional Theory (DFT) \cite{myrzakhmetov2024density}. Enhancing reaction conditions like light intensity, catalyst loading, and pH for maximum efficiency. Identifying and quantifying the role of defects in altering photocatalytic activity.

\begin{figure}[!ht]
    \centering
    \includegraphics[width=0.5\textwidth, angle=360]{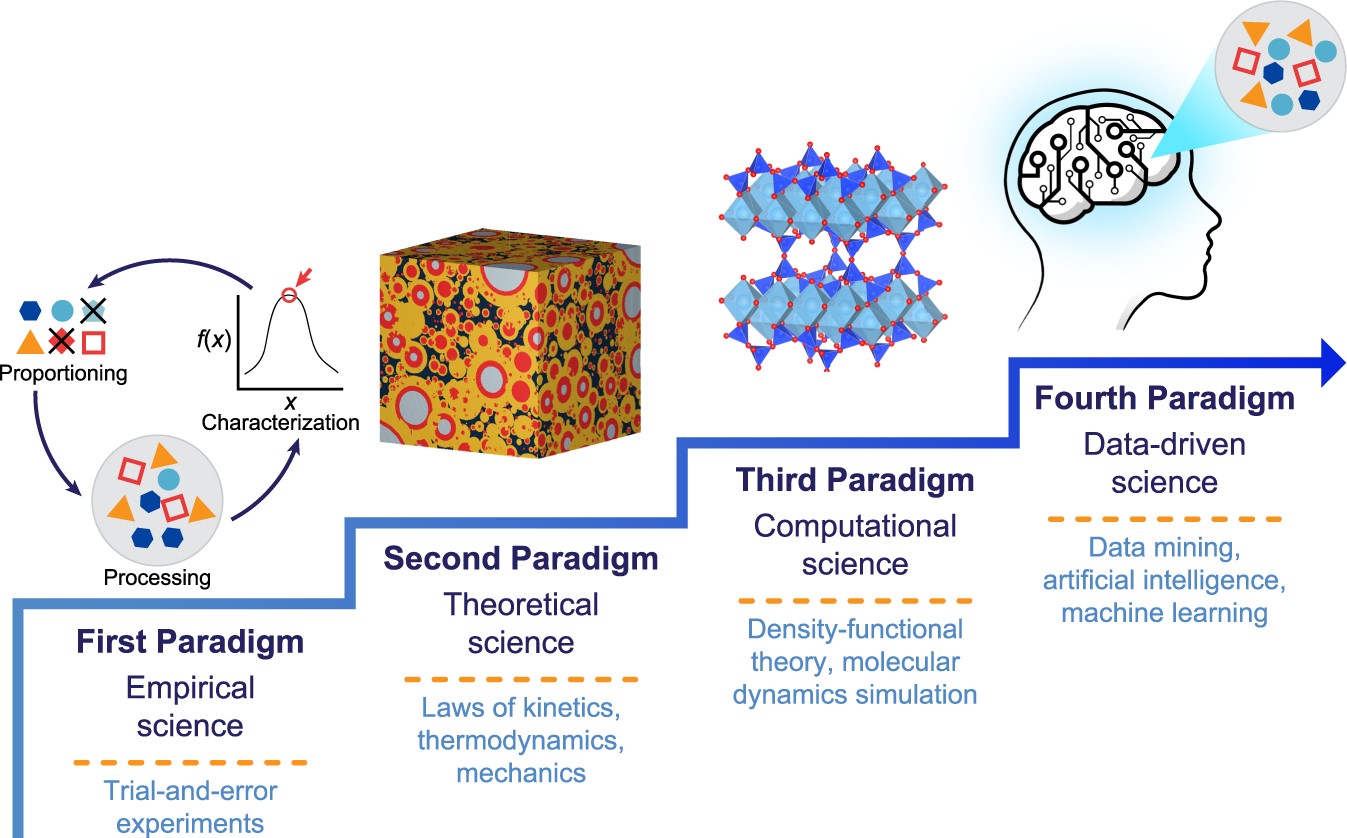}
    \caption{An illustrative computational paradigms for machine learning, adapted with permission from Li, 2022 \cite{li2022machine}}
    \label{fig11}
\end{figure} %open access article

%--------------------------------------------------------
\subsection{Classical Machine Learning in Material Design}
%--------------------------------------------------------
Several ML algorithms are now integral to photocatalytic research. Regression techniques are used to predict continuous variables such as bandgaps, absorption coefficients, and reaction rates. Linear regression are useful for establishing baseline relationships between material properties and photocatalytic performance. Support vector regression (SVR) \cite{awad2015support} are also effective for handling non-linear relationships in datasets, such as the influence of dopants on bandgap tuning. Gaussian process regression (GPR) \cite{deringer2021gaussian} are popularly known for its ability to quantify uncertainty in predictions, aiding in the exploration of unknown material spaces.

Classification algorithms \cite{kowsari2019text} analyze materials based on their ability to catalyze photoreactions. Random Forest (RF) \cite{rigatti2017random, breiman2001random} can deal with complex datasets and recognize important factors affecting activity. K-Nearest Neighbors (KNN) \cite{peterson2009k,al2024novel,javed2024evaluation} is helpful for grouping materials with comparable characteristics, supporting material evaluation. Decision Trees create understandable models for choosing the best photocatalysts.

Deep learning methods, especially neural networks, are frequently used in photocatalysis. Feedforward Neural Networks (FNNs) \cite{glorot2010understanding,yu2025construction,das2024artificial} can predict material characteristics and process effectiveness. Autoencoders can identify hidden features in datasets, allowing the discovery of new material combinations. Recurrent Neural Networks (RNNs) can analyze time-series data \cite{medsker2001recurrent,mienye2024recurrent,shapera2024recurrent}, such as reaction kinetics under changing conditions \cite{samuel2024machine}.

Generative models like Generative Adversarial Networks (GANs) \cite{mihandoost2024generative,yoon2024unveiling} and Variational Autoencoders (VAEs) \cite{pakornchote2024diffusion} are transforming the way we discover new materials. GANs can create hypothetical materials with desired characteristics, such as ideal bandgaps or high durability. VAEs, on the other hand, explore material design spaces by generating new structures based on existing data patterns.

Case studies demonstrate how GANs can be used to forecast materials with optimized bandgaps by learning from high-throughput DFT datasets \cite{hu2022high}. Additionally, VAEs help in designing structures that can withstand defects by predicting stable atomic arrangements.

Reinforcement learning is being used more and more to improve dynamic systems in photocatalysis. The algorithms find the best reaction conditions by constantly changing things like light strength, amount of catalyst, and temperature. Robotic systems using reinforcement learning can optimize synthesis processes, making them faster and cheaper.

Q-Learning \cite{watkins1992q,zhou2024optimized} shows how to efficiently adjust parameters to maximize the production of hydrogen through photocatalysis. Deep Deterministic Policy Gradient (DDPG) \cite{xie2024artificial} can handle continuous action spaces, making it useful for optimizing multi-variable reaction systems.

The application of machine learning (ML) in photocatalysis faces several challenges. Inconsistent experimental methods and measurement techniques lead to imprecise and incomplete datasets, which can be addressed by creating standardized datasets through collaborations and open data repositories. Many ML models operate as black boxes, making it hard to understand their predictions, but interpretable models like Random Forests or techniques like SHapley Additive exPlanations (SHAP) can help analyze the features. Experimental data is often limited due to the high costs of photocatalyst synthesis and testing, so transfer learning, where pre-trained models are adapted to specific tasks with limited data, can be a solution. Training complex models, such as deep neural networks, can be computationally intensive, but cloud computing and hybrid models can help balance accuracy and efficiency.

The future of machine learning in photocatalysis lies in its integration with experimental and computational methods. Automated high-throughput experimentation using robotic systems and machine learning algorithms can rapidly generate large datasets for model training. Combining data from diverse sources, such as spectroscopy, microscopy, and simulations, into unified machine learning frameworks can enable comprehensive material analysis. Transparent machine learning models can help bridge the gap between predictions and physical understanding. Federated learning approaches that leverage decentralized datasets from global research institutions can train robust models while maintaining data privacy.

%--------------------------------------------------------
\subsection{Quantum Machine Learning in Material Design} 
%--------------------------------------------------------
Quantum machine learning (QML) \cite{biamonte2017quantum,huang2021power} is an emerging field that combines quantum computing \cite{steane1998quantum} with machine learning techniques, offering unprecedented opportunities for accelerating material discovery and design \cite{jayan2024luminescent}. In the context of photocatalytic water splitting, QML has the potential to revolutionize the prediction and optimization of complex material properties by leveraging the unique computational advantages of quantum systems.

Quantum computing can simulate the quantum behavior of electrons in materials with exceptional precision. QML algorithms exploit this capability to efficiently navigate the vast design space of potential photocatalysts, identifying optimal configurations that traditional methods might miss. Classical machine learning models often struggle with capturing quantum mechanical phenomena such as electron correlation and spin effects. QML models, like quantum neural networks (QNNs) \cite{abbas2021power} and variational quantum circuits (VQCs) \cite{cerezo2021variational}, are explicitly designed to handle these phenomena, leading to more accurate predictions of electronic structures, bandgaps, and catalytic activities. High-throughput material screening is computationally expensive when using classical DFT calculations. QML can reduce the computational overhead by learning quantum behavior patterns from small datasets and extrapolating them to predict properties of unexplored materials as explouted by Ajagekar et al in Figure 12.

\begin{figure}[!ht]
    \centering
    \includegraphics[width=0.6\textwidth, angle=360]{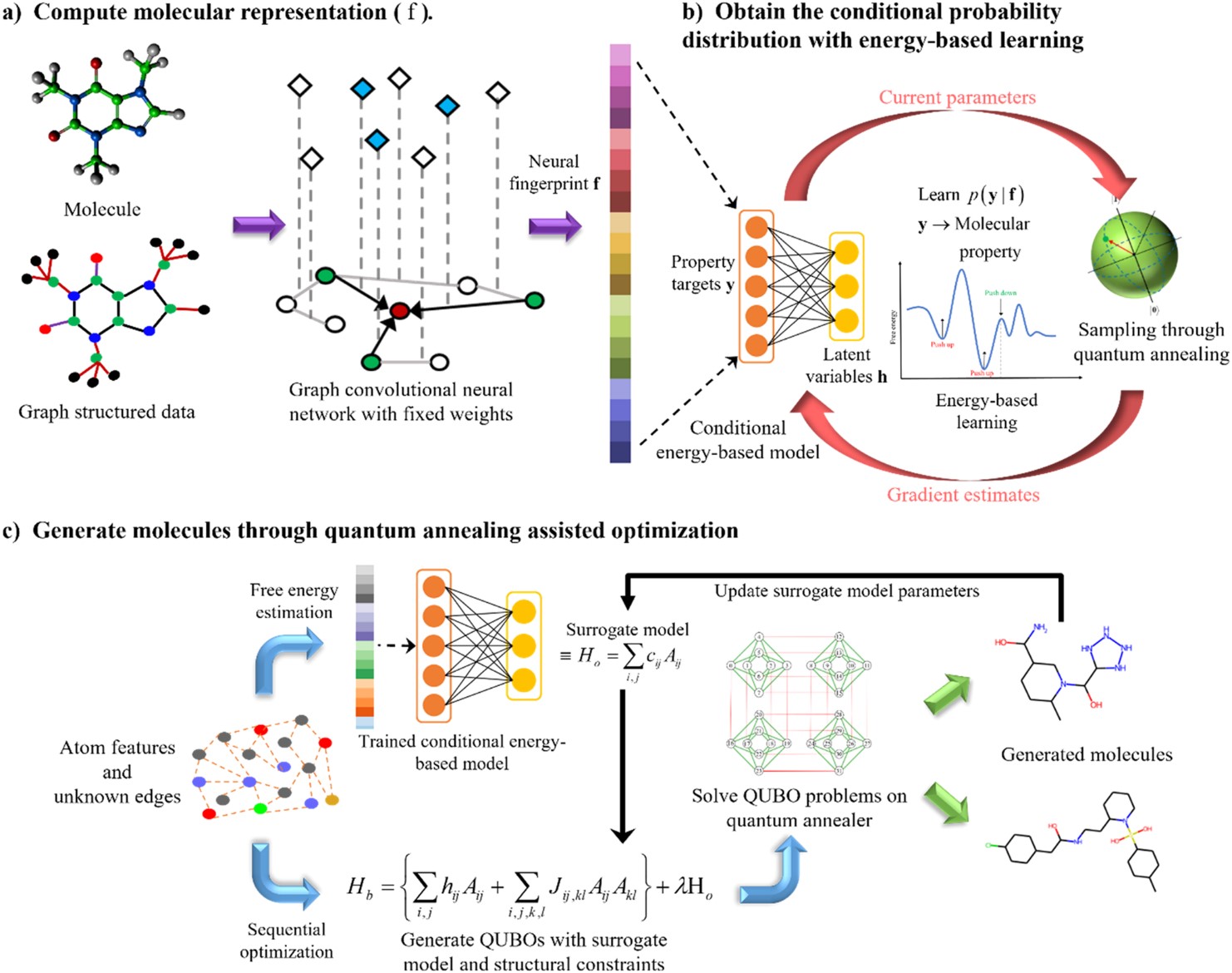}
    \caption{The energy-based model uses samples from a quantum annealer to link (a) molecular structures to their properties through a (b) GraphConv network. (c) It estimates free energy and optimizes molecular designs with desired properties by solving problems iteratively using the quantum annealer., adapted with permission from Ajagekar, 2023 \cite{ajagekar2023molecular}}
    \label{fig12}
\end{figure} %open access article

%--------------------------------------------------------
\subsection{Applications of QML in Material Design}
%--------------------------------------------------------
QML models can predict material properties like bandgaps and density of states more efficiently than traditional quantum chemistry methods \cite{huang2020quantum}. For instance, QML has been used to accelerate Hartree-Fock and DFT calculations in complex semiconducting systems, enabling faster discovery of materials with ideal light absorption and charge transport properties. By encoding chemical compositions and structural features into quantum representations, QML algorithms can explore non-linear relationships between material components and their photocatalytic performance. Defects play a critical role in tuning photocatalytic activity. QML models are adept at analyzing how atomic-scale defects influence electronic properties, offering new strategies for designing defect-engineered materials. Quantum systems can simultaneously optimize multiple objectives, such as maximizing light absorption and charge separation while minimizing recombination rates. This capability enables the design of materials tailored for specific photocatalytic applications.

Data scarcity is a common challenge in photocatalysis, as generating large datasets from experiments or simulations is costly. QML reduces the data requirements by efficiently utilizing quantum-enhanced algorithms that learn from limited data \cite{nandy2022audacity}. Photocatalysis involves complex interactions across electronic, structural, and thermal dimensions. QML leverages the exponential computational power of quantum systems to handle high-dimensional problems that are infeasible for classical ML models. By accurately simulating quantum behavior, QML serves as a bridge between theoretical predictions and experimental results, facilitating the validation and optimization of material models.

Quantum Support Vector Machines (QSVMs) use quantum kernels to capture the non-linear relationships in material datasets, providing robust classification of photocatalyst candidates \cite{rebentrost2014quantum,li2024quantum}. Quantum versions of GANs and VAEs have been applied to design hypothetical materials with optimal bandgap and stability properties. QRL has been used to optimize reaction pathways and synthesis conditions in photocatalytic systems by learning optimal strategies through quantum-enhanced exploration.

Combining quantum and classical computing models will likely be prevalent in the near future. These hybrid models use quantum processors for specific sub-problems, like quantum state evolution, while relying on classical processors for large-scale data processing. As quantum hardware improves, issues related to scalability and errors will decrease, allowing quantum machine learning to handle more complex material systems. Successful integration of quantum machine learning into material design requires collaboration between quantum physicists, chemists, and computer scientists, which will accelerate the development of quantum machine learning tools for practical applications in photocatalysis.

\begin{landscape}
\begin{longtable}{>{\hspace{0pt}}m{0.077\linewidth}>{\hspace{0pt}}m{0.034\linewidth}>{\hspace{0pt}}m{0.141\linewidth}>{\hspace{0pt}}m{0.113\linewidth}>{\hspace{0pt}}m{0.066\linewidth}>{\hspace{0pt}}m{0.089\linewidth}>{\hspace{0pt}}m{0.091\linewidth}>{\hspace{0pt}}m{0.12\linewidth}>{\hspace{0pt}}m{0.082\linewidth}}
\caption{Advances in DFT-AI integration for Photocatalytic Design for Water Splitting \label{tab:energies-1}}\\ 
\toprule
\textbf{Author(s)~}  & \textbf{Year~} & \textbf{DFT-AI }\par{}\textbf{Integration~}            & \textbf{Materials }\par{}\textbf{Investigated~} & \textbf{Light Source~} & \textbf{ML }\par{}\textbf{Architecture~} & \textbf{Dataset~}              & \textbf{Performance }\par{}\textbf{Metrics~} & \textbf{Software}\par{}\textbf{~(DFT/ML)~}  \endfirsthead 
\hline

Huo et al. \cite{huo2024feature} & 2024 & Feature-Assisted ML for band gap prediction & Binary semiconductors (1208) & Not specified & SVR, RF, GBDT, SISSO & High-throughput DFT (PBE functional) & RMSE \= 0.361 eV, R\(^2\) = 0.965 & VASP (DFT)/\par{} Scikit-learn (ML) \\
\hline

Wang et al. \cite{wang2024ml} & 2024 & ML-assisted screening for 2D materials & 2D materials (316,505) & Solar spectrum & ANN, RF, XGB & V2DB & RMSE \textless  0.4 eV (band gaps) & Quantum ESPRESSO/\par{} Python \\
\hline

Jyothirmai et al. \cite{jyothirmai2024machine} & 2024 & Catalyst Screening & g-C3N4/TMD Heterostructures & Not applicable & Random Forest Regression (RFR) & Calculated Gibbs Free Data & MAE & VASP (DFT)/\par{} Scikit-learn (ML)\\ 
\hline

Elbaz and Toroker \cite{elbaz2024density} & 2024 &  Bandgap Prediction &  Spinel Oxides (AB\(_2\)O\(_4\)) &  Not specified & Kernel Ridge Regression, SVR, RF & Simulated Spinel Dataset & RMSE: 0.02 eV (bandgap), Conductivity MAE: 5\% & NEGF \par{} GPAW/Scikit-learn \\ 
\hline

Moeini et al. \cite{sabagh2024machine} & 2024 & Hybrid ML for bandgap prediction & Low-symmetry perovskites & Not specified &  SVR, RFR, GBR, XGBoost & CMR dataset (1984 samples) & MAE \textless 0.1 eV for bandgap prediction & HSE06 /\par{} XGBoost \\ 
\hline

Oh et al. \cite{oh2024small} & 2024 & Small dataset ML for band engineering & ZnTe-based alloys & Solar spectrum & SISSO+$\alpha$-method & Custom ZnTe alloy database &RMSE: 0.1 eV (bandgap prediction) & Quantum ESPRESSO/\par{}SISSO\\ 
\hline

Liu et al. \cite{liu2022data} & 2022 & LightGBM for degradation rate prediction & Doped TiO\(_2\) (Ag, N, Cd) & UV-visible (254-600 nm) & LightGBM  & Experimental dataset (760 points) & R\(^2\) = 92.8\% & Jupyter/\par{}(Scikit-learn) \\ 
\hline

Kumar and Singh \cite{kumar2021chemical} & 2021 & Feature Engineering and SHAP & 2D octahedral materials (HfSe\(_2\), ZrSe\(_2\)) &  Visible light & Random Forest, SHAP & Open Quantum Materials DB & RMSE, R\(^2\) & VASP (DFT) /\par{}Scikit-learn (ML)\\ 
\hline

Gladkikh et al. \cite{gladkikh2020machine} & 2020 & Kernel Ridge Regression and Extremely Randomized Trees & ABX\(_3\) perovskites & Not specified & Kernel Ridge Regression & HSE06 bandgap data (199) & RMSE = 0.3 eV & GPAW/\par{} Python \\
\hline

Ren et al. \cite{ren2020machine} & 2020 & GPR and Sobol’ sensitivity analysis & TiO\(_2\) with Pt co-catalysts & Solar light simulation & Gaussian Process Regression & CFD and experimental & Yield optimization & CFD tools/\par{} Python (GPR) \\
\hline

Masood et al. \cite{masood2023enhancing} & 2019 & ML-accelerated DFT for material discovery & TiO\(_2\), CdS, WO\(_3\), g-C\(_3\)N\(_4\) & Visible light & Random Forest & Experimental and DFT outputs & Bandgap prediction (R\(^2\) = 0.95) & VASP /\par{} Scikit-learn\\ 
\hline

Wexler et al. \cite{wexler2018chemical} & 2018 & ML for HER descriptor discovery & Ni\(_2\)P with nonmetal dopants & Not specified & Regularized Random Forest & DFT-calculated structural data & HER $\Delta$ GH = $-0.11$ eV & Quantum ESPRESSO (DFT)/\par{} R (caret) \\ 
\hline

Li et al. \cite{li2018predicting} & 2018 & Predictive ML for Stability & Ternary Oxides & Not applicable & Kernel Ridge Regression, ET &ICSD (Crystal Structure DB) & RMSE, MAE & VASP (DFT)/\par{} PyCaret (ML) \\ 
\hline

Pereira et al. \cite{pereira2017machine} & 2017 & ML for HOMO and LUMO energy prediction & Organic molecules (111,725) & Not applicable & Random Forest & DFT-calculated HOMO/LUMO energies  & MAE = 0.15 eV (HOMO), 0.16 eV (LUMO) & GAMESS (DFT), R /\par{}  (RandomForest)\\
\hline

Hautier et al. \cite{hautier2010finding} & 2010 & Probabilistic Model & Ternary Oxides & Not applicable & Bayesian Modeling & ICSD & Stability Probability & Custom DFT Tools
\\
\bottomrule
\end{longtable}
\end{landscape}

%--------------------------------------------------------
\section{Convolutional Neural Network Applications}
%--------------------------------------------------------
Convolutional Neural Networks (CNNs) \cite{li2021survey} have revolutionized photocatalytic research, enabling precise analysis of material images, spectral data, and experimental results. Their ability to extract hierarchical features makes them invaluable for understanding and optimizing photocatalytic systems. A recent and promising development in this field is the application of Quantum Convolutional Neural Networks (QCNNs), which combine quantum computing’s power with the feature extraction capabilities of classical CNNs, providing unique advantages for solving complex material design challenges.

CNNs play a central role in automating the analysis of imaging, spectral, and reaction data. Their applications span from extracting morphological features from SEM, TEM, and AFM images, decoding patterns in UV-vis, XRD, Raman, and XPS spectra to predict material properties, and rapidly evaluating large datasets to identify promising photocatalysts.

%--------------------------------------------------------
\subsection{Quantum Convolutional Neural Networks (QCNNs)} 
%--------------------------------------------------------
Quantum Convolutional Neural Networks (QCNNs) \cite{cong2019quantum} leverage quantum computing to enhance the performance of classical CNNs. QCNNs utilize quantum states and quantum operations to process high-dimensional data more efficiently, offering advantages in speed, accuracy, and scalability for material science problems. QCNNs can process complex data, such as electron density maps or quantum states, directly, capturing features inaccessible to classical CNNs. For certain tasks, QCNNs outperform classical CNNs by exploiting the parallelism of quantum computing, especially when analyzing multi-modal datasets like combined imaging and spectroscopy data.

QCNNs are particularly useful in photocatalytic research for the following tasks: High-Dimensional Imaging Data: QCNNs excel at analyzing multi-dimensional datasets from advanced microscopy techniques like 3D TEM or tomography. Their ability to process quantum-mechanical data directly allows them to identify subtle features that classical CNNs might overlook.	Spectral Feature Recognition: QCNNs can identify non-linear patterns in XPS and UV-vis spectra more efficiently, aiding in the discovery of materials with desirable electronic properties. Bandgap Prediction: By combining quantum simulations and experimental data, QCNNs improve the prediction of electronic bandgaps, particularly for novel or hybrid materials. Defect and Interface Analysis: QCNNs analyze atomic-scale defects and heterojunction interfaces, offering insights into charge separation and recombination dynamics critical to photocatalytic performance.

Quantum Convolutional Neural Networks (QCNNs) have several benefits, they often achieve similar accuracy to classical Convolutional Neural Networks (CNNs) with less training data. Additionally, the quantum operations in QCNNs can provide insights into the physical connections between material properties and photocatalytic output. Furthermore, QCNNs can manage large datasets from high-throughput screening or multi-modal experiments more effectively. Current quantum computers have limitations in terms of noise and the number of available qubits, which affects the performance of Quantum Convolutional Neural Networks (QCNNs). Effectively combining QCNNs with traditional components for real-world use is still a challenge. Substantial research is needed to develop QCNN architectures optimized for specific tasks such as spectral forecasting or defect analysis.

%--------------------------------------------------------
\subsection{High-Throughput Workflows Using CNNs and QCNNs} 
%--------------------------------------------------------
Convolutional neural networks (CNNs) and quantum convolutional neural networks (QCNNs) are integrated into automated workflows to speed up the discovery of photocatalysts. Data processing pipelines use QCNNs to analyze raw data from imaging and spectroscopy, which provides insights to guide experimental designs. Robotic systems for synthesis and analysis utilize CNNs and QCNNs working together to optimize synthesis parameters in real-time based on experimental feedback. Predictive modeling combines QCNNs with machine learning models to predict material performance metrics from experimental and simulated data, enabling rapid material screening.

Limited availability of large, labeled datasets may hinder the use of both classical and quantum convolutional neural networks (CNNs and QCNNs). Decoding the decision-making process in these models remains a challenge, requiring the development of explainable AI tools. Combining QCNNs and classical CNNs can leverage the strengths of both models for efficient, scalable, and accurate analysis. Incorporating quantum mechanical principles into QCNN architectures as carefully exploited by Cong et al. in Figure 13 can enhance their predictive power for material science applications. Advancements in quantum hardware will enable QCNNs to handle larger datasets and perform more complex analyses.

\begin{figure}[!ht]
    \centering
    \includegraphics[width=0.6\textwidth, angle=360]{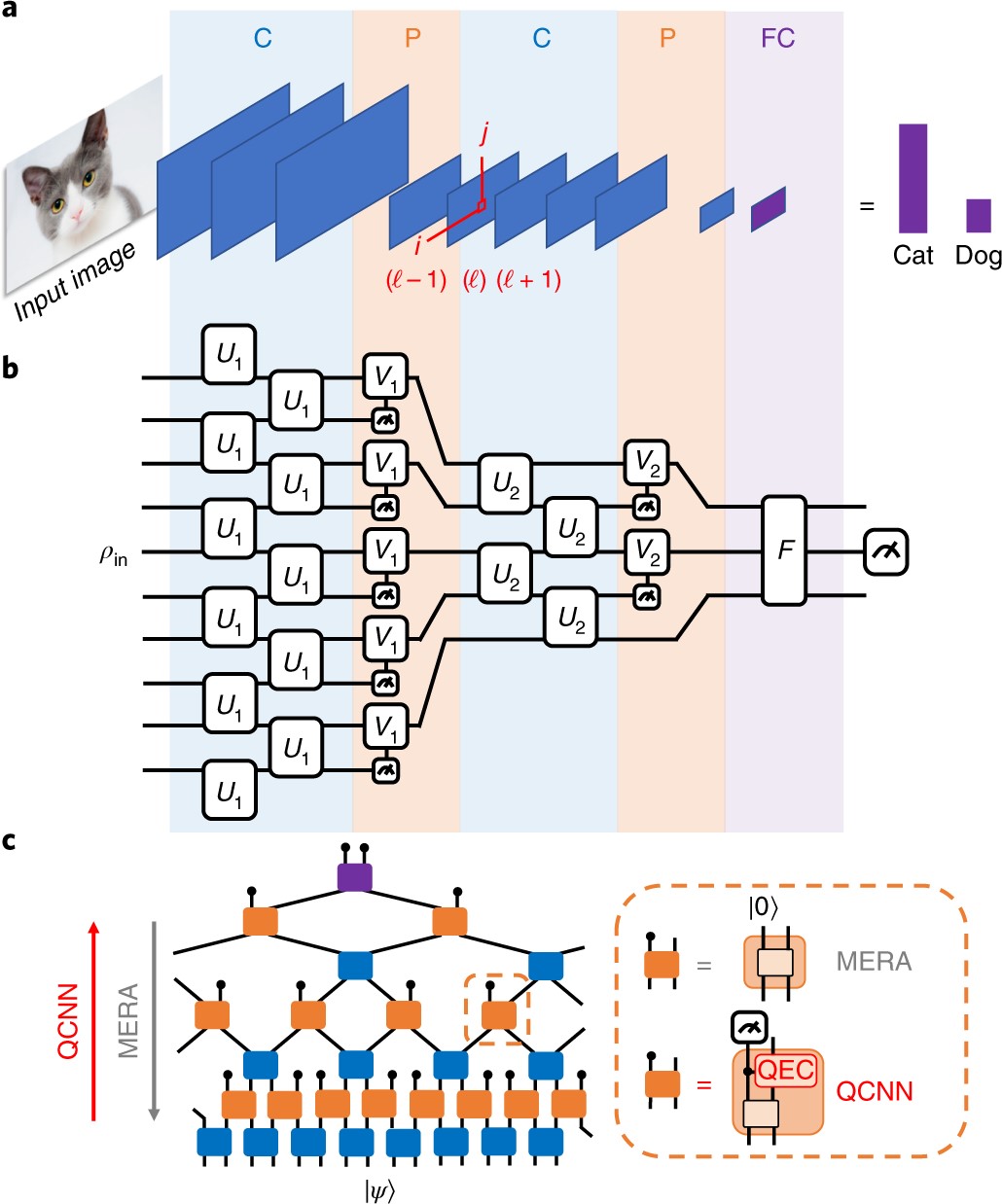}
    \caption{(a) Classical CNNs process images through layers to extract features and predict outcomes. (b) Quantum CNNs (QCNNs) use a similar layered structure with quantum gates and measurements. (c) QCNNs and MERA circuits share the same design but operate in opposite directions, adapted with permission from Cong, 2019 \cite{cong2019quantum}}
    \label{fig13}
\end{figure}

%--------------------------------------------------------
\section{DFT-AI Challenges and Future Perspective}
%--------------------------------------------------------
The integration of Density Functional Theory (DFT) with Artificial Intelligence (AI) has emerged as a powerful approach in accelerating the design and optimization of materials for photocatalytic water splitting. While this synergy offers numerous benefits, such as reducing computational costs and enabling high-throughput screening, it is not without challenges. This section examines the key challenges associated with the DFT-AI paradigm and explores future perspectives for advancing this integration in photocatalysis research.

%--------------------------------------------------------
\subsection{Challenges in DFT-AI Integration}
%--------------------------------------------------------
DFT calculations and experimental data often suffer from inconsistencies, making it difficult to train reliable AI models. Different research groups use varying exchange-correlation functionals, basis sets, and pseudopotentials, leading to non-uniform datasets. Inconsistencies in measurement techniques and environmental conditions can further complicate AI model training. Establish standardized protocols for DFT calculations and experimental validations. Initiatives like the Materials Project and Open Catalyst Project are paving the way by creating large, standardized databases.

While DFT is computationally efficient compared to other quantum mechanical methods, it remains resource-intensive for large-scale or high-throughput studies. Integrating AI models with DFT exacerbates this issue when training on vast datasets or performing iterative calculations. Use surrogate models like Gaussian Process Regression (GPR) or Neural Networks (NNs) to approximate DFT outputs for rapid predictions. Develop hybrid frameworks that combine low-fidelity models with high-fidelity DFT calculations to balance accuracy and efficiency.

AI models, particularly deep learning architectures, often function as black boxes, making it challenging to interpret their predictions and validate results against physical principles. Implement Explainable AI (XAI) techniques, such as SHapley Additive exPlanations (SHAP) or Local Interpretable Model-Agnostic Explanations (LIME), to identify key features driving predictions. Use physics-informed neural networks (PINNs) to incorporate known physical laws directly into AI models.

AI models trained on specific material systems often struggle to generalize to new compositions or structures, limiting their utility in discovering novel photocatalysts. Employ transfer learning, where pre-trained AI models are fine-tuned for specific material classes with limited data. Use unsupervised learning methods to cluster similar materials and identify transferable patterns.

%--------------------------------------------------------
\subsection{Future Perspectives in DFT-AI Integration}
%--------------------------------------------------------
Integrating quantum mechanics into AI models represents a significant frontier in improving the accuracy and scalability of DFT-AI systems. Quantum Machine Learning (QML) leverages quantum computing to accelerate the training and inference of AI models, particularly for DFT-like simulations, by processing complex quantum states more efficiently than classical methods. Quantum Neural Networks (QNNs) combine the strengths of quantum systems and AI algorithms, offering powerful tools to solve high-dimensional problems in material science, such as the prediction of electronic structures and reaction dynamics. These quantum-inspired approaches promise to reduce computational costs and unlock novel pathways for designing advanced photocatalysts.

Automated AI-driven workflows are also poised to revolutionize material discovery by enabling inverse design and active learning frameworks. Inverse design uses AI to propose material structures or compositions with desired properties, which are then validated through DFT calculations. Active learning further enhances efficiency by iteratively selecting the most informative data points, optimizing computational and experimental resources. Multi-fidelity modeling provides additional scalability by combining low-cost approximations, such as semi-empirical methods, with high-accuracy DFT calculations. This hybrid approach allows researchers to generate large training datasets for AI models while maintaining precision, bridging the gap between speed and computational intensity.

Finally, expanding data repositories and integrating multi-modal data are essential for advancing DFT-AI integration. Initiatives like the Materials Project, JARVIS-DFT, and Open Catalyst Project provide standardized datasets, but future repositories must also incorporate multi-modal data, such as electronic structures, optical properties, and experimental synthesis conditions. Graph Neural Networks (GNNs) offer a promising solution for representing atomic structures as graphs, enabling efficient learning of material-property relationships. Multi-task learning can further enhance material screening by simultaneously predicting multiple properties, such as bandgap, stability, and defect tolerance. These advancements will facilitate a more comprehensive understanding of photocatalyst performance, accelerating innovation in sustainable energy solutions.

%--------------------------------------------------------
\section{Conclusion}
%--------------------------------------------------------
The pursuit of sustainable hydrogen production through photocatalytic water splitting has seen remarkable progress, driven by advancements in materials science, computational techniques, and artificial intelligence. This review has explored the intricate interplay between light-driven photoreactions, semiconductor materials, reaction mechanisms, machine learning algorithms, and the integration of Density Functional Theory (DFT) with AI. Together, these innovations underscore the transformative potential of photocatalytic systems to reshape the global energy landscape while addressing urgent environmental challenges.

The evolution of semiconductors has expanded the scope of photocatalysis, transitioning from traditional materials like TiO\(_2\) and ZnO to advanced systems such as hybrid quantum dots and perovskites. Upconversion materials and tandem systems have improved solar spectrum utilization, enhancing light absorption and reaction efficiency. A deeper understanding of photon absorption, charge carrier dynamics, and surface reactions has facilitated the design of highly efficient water-splitting systems. Innovations like plasmonic enhancement, Z-scheme heterojunctions, and dual-functional catalysts have addressed critical challenges such as charge recombination and stability, further advancing the field.

Artificial intelligence, particularly machine learning, has revolutionized photocatalyst discovery by enabling high-throughput screening, optimizing reaction conditions, and predicting material properties. Techniques like Quantum Machine Learning (QML) and Quantum Convolutional Neural Networks (QCNNs) have shown promise in solving high-dimensional problems and accelerating material design. Similarly, the synergy between DFT and AI has streamlined the prediction of electronic, structural, and optical properties, though challenges like data quality, computational costs, and model interpretability persist.

Despite these advancements, unresolved challenges remain in achieving scalable and cost-effective photocatalytic hydrogen production. Material stability continues to be a significant hurdle, as many promising photocatalysts degrade under reaction conditions. Efficiency improvements, particularly in solar-to-hydrogen (STH) conversion rates, demand ongoing innovation in light harvesting and charge separation techniques. The lack of standardized datasets for AI training and high costs associated with certain materials and computational methods also hinder scalability and reproducibility. Addressing these barriers is essential for the commercial adoption of photocatalytic technologies.

Future research directions include the development of hybrid photocatalysts that combine organic and inorganic components for improved stability and efficiency, as well as leveraging defect engineering and surface functionalization to optimize catalytic activity. Quantum technologies, such as quantum computing and QML, offer transformative potential for simulating complex material systems and optimizing reaction pathways. Automated and high-throughput approaches, including robotic synthesis systems and AI-enabled real-time reaction monitoring, will further accelerate the discovery and optimization of photocatalysts. Additionally, fostering interdisciplinary collaborations among chemists, physicists, material scientists, and AI researchers, along with the creation of global databases and open-access repositories, will enhance data sharing and accelerate progress.

The field of photocatalytic water splitting stands at the nexus of materials science, computational modeling, and AI, offering a promising pathway toward sustainable hydrogen production. While challenges persist, the rapid pace of innovation suggests that scalable solutions are within reach. By fostering interdisciplinary collaboration and leveraging emerging technologies, the scientific community can unlock the full potential of photocatalytic systems, paving the way for a cleaner and more sustainable energy future.

%--------------------------------------------------------
\section{Acknowledgment}
%--------------------------------------------------------
This research was not funded. Any opinions, findings, conclusions, or recommendations expressed in this review are those of the author(s) and do not necessarily reflect the views of their respective affiliations.

%--------------------------------------------------------
\section{Conflicts of interest}
%--------------------------------------------------------
The author(s) declare no competing interests.

\nomenclature{KS}{Kohn-Sham}
\nomenclature{AI}{Artificial Intelligence}
\nomenclature{AEM}{Anion Exchange Membrance}
\nomenclature{BSD}{Berkeley Software/Standard Distribution}
\nomenclature{CUDA}{Compute Unified Device Architecture}
\nomenclature{CNN}{Convolutional Neural Network}
\nomenclature{CPU}{Central Processing Unit}
\nomenclature{DAE}{Direct Air Electrolysis}
\nomenclature{DFT}{Density Functional Theory} 
\nomenclature{DL}{Deep Learning}
\nomenclature{DOE}{Department of Energy}
\nomenclature{ECL}{Educational Community License}
\nomenclature{GGA}{Generalized Gradient Approximation}
\nomenclature{GHG}{Greenhouse gases}
\nomenclature{GPL}{General Public License}
\nomenclature{GPU}{Graphical Processing Unit}
\nomenclature{LDA}{Local Density Approximation}
\nomenclature{LGPL}{GNU Lesser General Public License} 
\nomenclature{LTH}{Light-to-Hydrogen}
\nomenclature{MEC}{Microbial Electrolysis Cells}
\nomenclature{MIT}{Massachusetts Institute of Technology}
\nomenclature{MD}{Molecular Dynamics}
\nomenclature{ML}{Machine Learning }
\nomenclature{PBE0}{Modified Perdew–Burke–Ernzerhof}
\nomenclature{PEC}{Photo-electrochemical Catalysis}
\nomenclature{PEM}{Proton Exchange Membrane}
\nomenclature{STH}{Solar-to-Hydrogen}
\nomenclature{SciAI}{Science and Artificial Intelligence}
\nomenclature{XC}{Exchange-Correlation}
\nomenclature{AIMD}{Ab Initio Molecular Dynamics} 
\nomenclature{HSE03}{Heyd-Scuseria-Ernz-Erhof } 
\nomenclature{TDDFT}{time-dependent density functional theory} 
\nomenclature{ASAP}{Atomistic Simulation Advanced Platform} 
\nomenclature{NIR}{near-infrared} 
\nomenclature{LEDs}{light-emitting diodes}
\nomenclature{STH}{solar-to-hydrogen} 
\nomenclature{UCNPs}{Upconversion nanoparticles} 
\nomenclature{HER}{hydrogen evolution reaction} 
\nomenclature{OER}{oxygen evolution reaction} 
\nomenclature{LSPR}{localized surface plasmon resonance}
\nomenclature{HQDs}{Hybrid quantum dots} 
\nomenclature{ETU}{energy transfer upconversion} 
\nomenclature{VB}{valence band} 
\nomenclature{CB}{conduction band} 
\nomenclature{SVR}{Support vector regression} 
\nomenclature{RF}{Random Forest} 
\nomenclature{KNN}{K-Nearest Neighbors} 
\nomenclature{FNNs}{Feedforward Neural Networks} 
\nomenclature{RNNs}{Recurrent Neural Networks} 
\nomenclature{GANs}{Generative Adversarial Networks} 
\nomenclature{VAEs}{Variational Autoencoders}
\nomenclature{DDPG}{Deep Deterministic Policy Gradient} 
\nomenclature{QML}{Quantum machine learning} 
\nomenclature{QNNs}{quantum neural networks} 
\nomenclature{VQCs}{variational quantum circuits} 
\nomenclature{QSVMs}{Quantum Support Vector Machines} 
\nomenclature{QCNNs}{Quantum Convolutional Neural Networks} 
\nomenclature{GPR}{Gaussian Process Regression} 
\nomenclature{NNs}{Neural Networks} 
\nomenclature{XAI}{Implement Explainable AI} 
\nomenclature{SHAP}{SHapley Additive exPlanations} 
\nomenclature{LIME}{Local Interpretable Model-Agnostic Explanations} 
\nomenclature{PINNs}{physics-informed neural networks} 
\nomenclature{SIESTA}{Spanish Initiative for Electronic Simulations with Thousands of Atoms} 
\nomenclature{VASP}{Vienna Ab initio Simulation Package} 
\nomenclature{ASAP}{Atomistic Simulation Advanced Platform} 
\nomenclature{CASTEP}{CAmbridge Serial Total Energy Package.} 
\nomenclature{SEM}{Scanning electron microscope} 
\nomenclature{TEM}{Transmission Electron Microscopy} 
\nomenclature{AFM}{Atomic force microscopy} 
\nomenclature{UV-vis}{Ultraviolet–visible spectroscopy} 
\nomenclature{XRD}{X-ray crystallography} 
\nomenclature{XPS}{X-ray photoelectron spectroscopy} 
\nomenclature{GBDT}{Gradient Boosted Decision Trees} 
\nomenclature{SISSO}{Sure independence screening and sparsifying operation} 
\nomenclature{XGB}{Extreme Gradient Boosting} 
\nomenclature{RFR}{Random Forest Regression} 
\nomenclature{LightGBM}{Light Gradient-Boosting Machine} 
\nomenclature{CFD}{Computational Fluid Dynamics} 
\nomenclature{ICSD}{Inorganic Crystal Structure Database} 
\nomenclature{CMR}{Cardiovascular magnetic resonance} 
\nomenclature{V2DB}{BridgeData V2} 
\nomenclature{HOMO}{Highest Occupied Molecular Orbital} 
\nomenclature{LUMO}{Lowest Unoccupied Molecular Orbital} 
\printnomenclature

%%%%%%%%%%%%%%%%%%%%%%%%%%%%%%%%%%%%%%%%%%%%%%%%%%%%%%%%%%%%%%%%%%%%%%%%

%% The Appendices part is started with the command \appendix;
%% appendix sections are then done as normal sections
%\appendix

%\section{Sample Appendix Section}
%\label{sec:sample:appendix}
%Lorem ipsum dolor sit amet, consectetur adipiscing elit, sed do eiusmod tempor section \ref{sec:sample1} incididunt ut labore et dolore magna aliqua. Ut enim ad minim veniam, quis nostrud exercitation ullamco laboris nisi ut aliquip ex ea commodo consequat. Duis aute irure dolor in reprehenderit in voluptate velit esse cillum dolore eu fugiat nulla pariatur. Excepteur sint occaecat cupidatat non proident, sunt in culpa qui officia deserunt mollit anim id est laborum.

%% If you have bibdatabase file and want bibtex to generate the
%% bibitems, please use
%%
\bibliographystyle{elsarticle-num} 
\bibliography{cas-refs}

%% else use the following coding to input the bibitems directly in the
%% TeX file.

 %\begin{thebibliography}{00}

%\bibitem{label}
% %% Text of bibliographic item

% \bibitem{}

 %\end{thebibliography}
\end{document}